%% file: main.tex
\documentclass[graybox]{svmult}

\usepackage{type1cm}        
\usepackage{makeidx}         
\usepackage{graphicx}        
\usepackage{multicol}        
\usepackage[bottom]{footmisc}
\usepackage{paralist}
\usepackage{amsmath}
\usepackage{comment}
\usepackage{mathtools,algpseudocode,algorithm}
\usepackage{tikz} 
\usepackage{pdflscape}
\usepackage{newtxtext}       %
\usepackage[varvw]{newtxmath}       

\usepackage[caption=false]{subfig}
\usepackage[export]{adjustbox}

\usepackage{pgfplots}
  \pgfplotsset{compat=newest}
  \usetikzlibrary{plotmarks}
  \usetikzlibrary{arrows.meta}
  \usepgfplotslibrary{patchplots}

\makeindex             

\usepackage{acronym}
\usepackage{titlesec}

\definecolor{matlab_blue}{rgb}{0.00,0.00,1.00}
\newcommand{\bluecross}{\tikz[baseline]{\draw[matlab_blue,solid,line width = 1pt](0mm,0mm) -- (1.5mm,1.5mm) -- (0.75mm,0.75mm) -- (1.5mm,0mm) -- (0mm,1.5mm)}}
\definecolor{matlab_black}{rgb}{0.00,0.00,0.00}
 \newcommand{\blackcircle}{\tikz[baseline]{\draw[black,fill,solid,line width = 0.5pt] (0,0.8mm) circle (0.1)}}
\definecolor{matlab_red}{rgb}{1.00,0.00,0.00}
\newcommand{\redsquare}{\tikz[baseline]{\draw[matlab_red,fill,solid,line width = 1pt](0mm,0mm) -- (1.5mm,0) -- (1.5mm,1.5mm) -- (0mm,1.5mm) -- (0mm,0mm)}}
\definecolor{matlab_orange}{rgb}{0.87,0.49,0.00}
\newcommand{\orangesquare}{\tikz[baseline]{\draw[matlab_orange,solid,line width = 1pt](0mm,0mm) -- (1.5mm,0) -- (1.5mm,1.5mm) -- (0mm,1.5mm) -- (0mm,0mm)}}

\titlespacing\section{0pt}{14pt plus 4pt minus 2pt}{14pt plus 2pt minus 2pt}
\titlespacing\subsection{0pt}{12pt plus 4pt minus 2pt}{12pt plus 2pt minus 2pt}

\input{acronyms.tex} 
\newcommand{\todo}[1]{\textcolor{red}{\footnotesize{\textsf {[todo: #1]}}}}

\newcommand{\instr}[1]{\textcolor{blue}{\footnotesize{\textsf {[instructions: #1]}}}}
\begin{document}

\title*{MmWave Mapping and SLAM for 5G and Beyond}
\author{Yu Ge, Ossi Kaltiokallio, Hyowon Kim, Jukka Talvitie,  Sunwoo Kim, Lennart Svensson, Mikko Valkama, Henk Wymeersch}
\institute{Yu Ge, Hyowon Kim, Lennart Svensson, Henk Wymeersch \at Chalmers University of Technology, 41296 Gothenburg, Sweden \\ \email{yuge,hyowon,lennart.svensson,henkw@chalmers.se}
\and Ossi Kaltiokallio, Jukka Talvitie, Mikko Valkama \at Tampere University,  33101 Tampere, Finland \\ \email{ ossi.kaltiokallio,jukka.talvitie,mikko.valkama@tuni.fi}
\and 
Sunwoo Kim \at Hanyang University, Seoul, 04763, South Korea 
\email{remero@hanyang.ac.kr}
}
%
%
\maketitle

\abstract*{The sensing aspect in integrated sensing and communication is provided through either mono-static or bi-static sensing. Mono-static sensing is most commonly associated with radar-like mapping, where a sensor maps an environment and tracks moving objects in a local reference frame. In contrast bi-static mapping is most commonly associated with positioning / localization, where a user device 3D position is determined in a global reference frame, while environment mapping is mainly a side-product used to improve localization accuracy and coverage. In this chapter, we will provide an overview of the fundamental tools used to solve mapping, tracking, and simultaneous localization and mapping problems. In particular, methods based on random finite set theory and Bayesian graphical models will be introduced. A numerical study will compare these approaches in a variety of scenarios and contrast 5G mmWave with 6G D-band performance. The use of intelligent surfaces will also be considered.}

\abstract{
Device localization and radar-like mapping are at the heart of integrated sensing and communication, enabling not only new services and applications, but can also improve communication quality with reduced overheads. These forms of sensing are however susceptible to data association problems, due to the unknown relation between measurements and detected objects or targets. 
In this chapter, we  provide an overview of the fundamental tools used to solve mapping, tracking, and simultaneous localization and mapping (SLAM) problems.
We distinguish the different types of sensing problems and then focus on mapping and SLAM as running examples. Starting from the applicable models and definitions, we describe the different algorithmic approaches, with  a particular focus on how to deal with data association problems. 
In particular, methods based on random finite set theory and Bayesian graphical models are introduced in detail. A numerical study with synthetic and experimental data is then used to compare these approaches in a variety of scenarios. }


\section{Motivation and Introduction}
\Ac{ISAC}  has become one of the main differentiators of 6G with respect to previous generations of communication systems \cite{liu2022integrated}. Foreseen applications of \ac{ISAC} in 6G include providing useful information for optimizing communication metrics, as well as 
the support of challenging use cases, such as extended reality (XR) and cooperating robots, which require high data rate communication and cm-level, ultra-fast sensing. Opening up for operation at higher frequencies (such as D-band between 130 GHz and 170 GHz) where more bandwidth is available for Gb/s data rates, complemented with highly directional transmission with large aperture arrays, will lead to 
high delay and angle resolution, needed to provide accurate 6D (3D orientation and 3D position) localization and sensing. The high carrier frequencies also lead to new opportunities for material characterization and spectroscopy. While it is yet unclear which of these applications will be part of the 6G ecosystem, the widespread use of sensing is unquestionable. Already in 5G mmWave, sensing can play an important role, providing significant situational awareness with little or no change to existing signals and infrastructure.

\begin{figure}
    \centering
    \includegraphics[width=\textwidth]{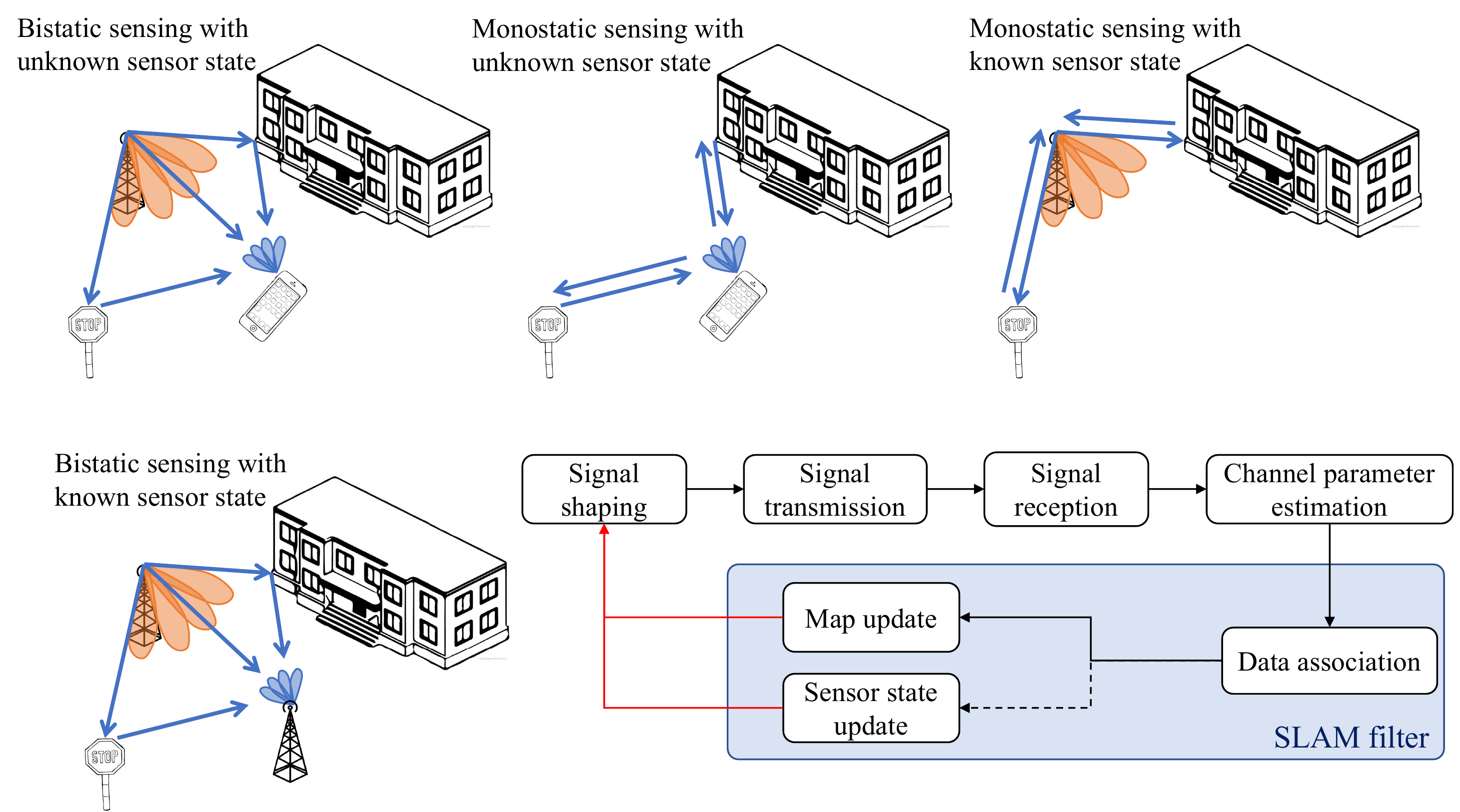}
    \caption{Examples of different sensing modalities. In the flowchart, the main steps in radio SLAM are shown, starting with optimized signal shaping, based on prior sensor and map information, followed by signal transmission (either remote or at a sensor) and reception. Then a channel estimation routine is executed, which provides range and angle measurements. These measurements must be associated with previously seen targets, before the map and sensor state are updated.}

    \label{fig:overview}\vspace{-4mm}
\end{figure}

The term \emph{sensing} is often limited from its broader  definition (i.e., to detect events, to measure changes in the environment, or to measure a physical property) to mean radar-like sensing. However, it can also cover uplink and downlink channel (parameter) estimation, radio frequency sensing, spectroscopy, weather monitoring, as well as any downstream processes that rely on sensing data. In this sense, 5G and earlier generations already perform some form of sensing, if only for channel estimation \cite{berger2010application} and positioning  \cite{dwivedi2021positioning}. This definition of sensing also implies that \ac{ISAC} is much more than simply using a common waveform over a common hardware and processing the backscattered signal, and thus also includes localizing connected \acp{ue} and improving communication by aid of sensing information. This broader definition thus accounts for a large potential for 6G systems, a potential that can already be applied to 5G, but is largely untapped.  
Sensing can be roughly grouped into 3 categories: monostatic, bistatic, and multistatic sensing. In monostatic sensing, the transmitter and receiver are co-located and thus share complete knowledge of the transmitted signal and clock \cite{sturm2011waveform}. This type of sensing is, for example, used in automotive radar, where static landmarks are mapped and/or dynamic objects are tracked. In a communication system, monostatic sensing can rely on abundant data signals, which improves availability and accuracy compared to the use of sparse pilots. In bistatic sensing, the receiver may only have partial knowledge of the transmitted signal and may not have access to the transmitter clock.  Finally, in multistatic sensing, several receivers are employed to perform the sensing task. In certain sensing problems, the receiver or transmitter may itself have an unknown state (e.g., position and orientation), which must be inferred while tracking objects \cite{kim20205g}. An example is radio localization, which relies on dedicated pilot signals to determine a \ac{ue}'s position and clock bias \cite{dwivedi2021positioning}.
The transmitter or receiver with unknown state is (confusingly) called the \emph{sensor} in the tracking and mapping literature. Solving problems with unknown sensor state include \ac{SLAM} \cite{mullane2011random} (when the sensed objects/targets are static landmarks) and simultaneous localization and tracking (SLAT) \cite{meyer2018message} (when the sensed objects/targets are themselves moving). An overview of the different sensing modalities is depicted in Fig.~\ref{fig:overview}.

Nearly all sensing applications that involve multiple objects/targets suffer from a so-called \ac{DA} problem, though it is manifested in different ways for different types of sensing tasks \cite{mahler2004statistics}. In monostatic sensing, DA occurs due to the need to relate measurement at one time to detected objects (or tracks) at a previous time. In multistatic sensing, DA occurs due to the unknown relation between measurements at one receiver to measurements at another receiver, corresponding to the same object. In bistatic sensing, DA occurs similar to monostatic sensing due to the unknown relation between measurements and previously detected objects or tracks. Hence, in monostatic and bistatic sensing, DA is due to tracking over time, while snapshot-based DA is only present in multistatic sensing.
To solve DA problems, there exist a variety of solutions, ranging from the classic linear assignment algorithms to find the best association to probabilistic methods based on \ac{BP} or \ac{RFS} theory. Table \ref{tab:classification} provides an overview of the different mapping and tracking problems. Note that the pure localization problem (e.g., as in GPS or in 5G) does not suffer from any DA problem, since the different sources use distinct and separable signals. 
\vspace{-5mm}

\begin{table}
\caption{{Problem classification.}}
\centering
    \begin{tabular}{llll}
    \hline
    \textbf{Term} & \textbf{Objects} & \textbf{Sensor state} & \textbf{Data association}\\
    \hline 
    mapping & static (landmarks) & known & unknown\\
    tracking & moving (targets) & known & unknown\\
    localization & ~~-- & unknown & known\\
    SLAM & static (landmarks) & unknown & unknown\\
    SLAT & moving (targets) & unknown & unknown\\
    \hline 
    \end{tabular}
    \label{tab:classification}
\end{table}

Given the importance of solving DA problems in \ac{ISAC} in all types of sensing tasks, our ambition with this chapter is two-fold. First of all, we aim to give the reader with an introduction to mapping, tracking, localization, and SLAM in the context of mmWave communication, considering both monostatic and bistatic sensing. Second, we aim to provide a gentle introduction to the fundamentals of \ac{RFS} theory filters to solve the corresponding DA problems. We  note that mapping and tracking are special cases of SLAM and SLAT, respectively. For that reason, our focus will be on SLAM, while the corresponding mapping approaches are then easily obtained by fixing the sensor state.

\runinhead{Chapter organization} 
The remainder of this chapter is structured as follows. In Sect.~\ref{sec:scenarios}, we describe a scenario for SLAM, considering both monostatic and bistatic settings. Important concepts such as state, dynamics, and measurements are introduced, along with the associated notations. In Sect.~\ref{sec:methods}, the different methodologies for solving the \ac{ISAC} SLAM problems are covered, starting from a broad introduction of the field, and then zooming in to two Bayesian approaches: one using \ac{RFS} theory and another one using \ac{BP}. Without any intention for mathematical completeness, several practical SLAM methods are described at a high level. Exemplary results using simulated and experimental data are provided in Sect.~\ref{sec:results}. Finally, Sect.~\ref{sec:outlook} concludes this chapter with an outlook of the main problems we foresee and potential solution strategies. These unsolved problems give an indication of the richness of \ac{ISAC} SLAM.

\section{Scenarios and Models} \label{sec:scenarios}

\begin{figure}
    \centering
    \includegraphics[width=0.7\textwidth]{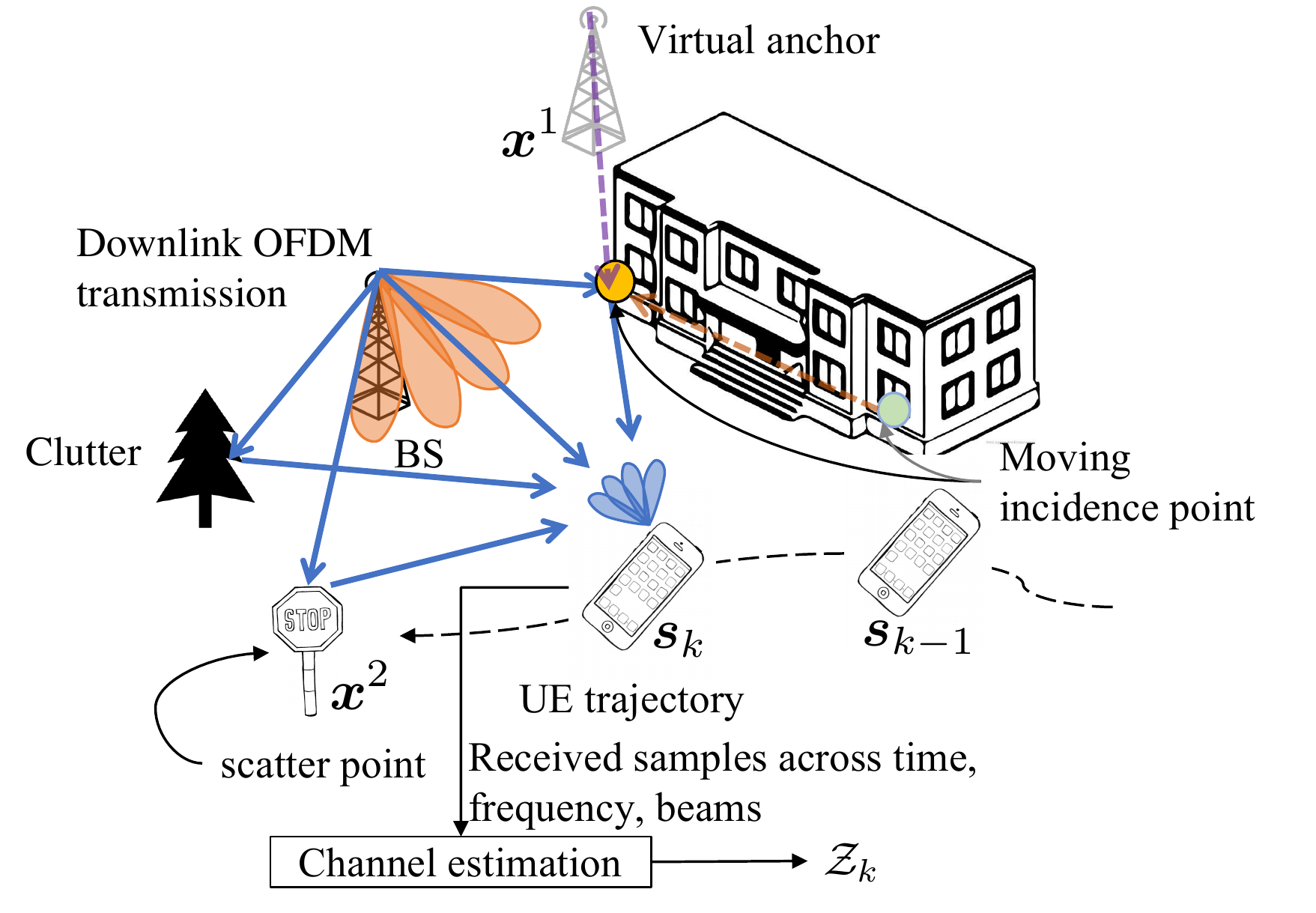}
    \caption{A mmWave downlink scenario with a single \ac{bs} and a moving \ac{ue}. The location of the \ac{bs} is known. The BS sends \acf{OFDM} signals to the \ac{ue} over the environment. The \ac{ue} utilizes the received signals to estimate a set of channel parameters $\mathcal{Z}_k$, which depend on the underlying geometry and can be used by the \ac{ue} to track its own state and build a map of the environment.}
    \label{fig:scenario}\vspace{-4mm}
\end{figure}

The considered scenario is illustrated in Fig.~\ref{fig:scenario} in which a single \ac{bs} periodically sends a downlink \ac{prs} to the \ac{ue} and the signal can reflect and scatter as it propagates through the medium. The physical structures in the environment that reflect and scatter the signals can be decomposed into parametric point representation referred to as landmarks. Reflecting surfaces are modeled by the mirror image of the \ac{bs}, as \acp{va}. Small objects that scatter the signal are modeled by their location, as \acp{sp} \cite{kim20205g}. The state of the $i$-th landmark is $\boldsymbol{x}^i = [(\boldsymbol{x}^i_\textrm{LM})^\top \; m^i]^\top$, where $\boldsymbol{x}^i_\textrm{LM} \in \mathbb{R}^3$ represents the position of landmark $i$ and $m^i \in \lbrace BS, VA,SP \rbrace$ represents the landmark type. If the environment consists of $I$ landmarks in total, the map of the environment can be represented by a set of all landmarks $\mathcal{X} = \lbrace \boldsymbol{x}^1, \ldots, \boldsymbol{x}^I \rbrace$.


The full \ac{ue} state comprises of the pose and clock. The pose represents the position and orientation of the \ac{ue}, whereas the clock represents the required parameters needed to synchronize the local clock to the network clock (perfect synchronization of the \ac{ue} to the \ac{bs} is not a reasonable assumption in practice). Let $\boldsymbol{s}_{k-1}$ denote the \ac{ue} state at time $k-1$ and assuming the process noise is zero-mean Gaussian, the transition density of the \ac{ue} state can be expressed as
\begin{equation}\label{transition_density}
    f(\boldsymbol{s}_{k}|\boldsymbol{s}_{k-1},\boldsymbol{u}_{k})=\mathcal{N}(\boldsymbol{s}_{k};\boldsymbol{v}\left(\boldsymbol{s}_{k-1},\boldsymbol{u}_{k} \right),\boldsymbol{Q}_{k-1}),
\end{equation}
where $\boldsymbol{Q}_{k-1}$ is the process noise covariance and $\boldsymbol{v}(\cdot)$ the transition model in which $\boldsymbol{u}_{k}$ denotes a known control input. 

At time step $k$, the \ac{ue} receives \ac{OFDM} downlink signals from the \ac{bs}, e.g., as modeling in \cite{heath2016overview}, where the propagation channel over subcarrier $\kappa$ can be expressed as 
\begin{align}
    \boldsymbol{H}_{\kappa,k}=\sum_{i=1}^{I_k} g_{k}^{i} e^{-\jmath 2 \pi \kappa \Delta_f \tau^{i}_k} \boldsymbol{a}_{\text{UE}}(\boldsymbol{\theta}_{k}^{i})\boldsymbol{a}^\top_{\text{BS}}(\boldsymbol{\phi}_{k}^{i}),
\end{align}
where $\Delta_f$ is the subcarrier spacing, and $\boldsymbol{a}_{\text{UE}}(\cdot)$ and $\boldsymbol{a}_{\text{BS}}(\cdot)$ are the \ac{ue} and \ac{bs} response vectors. 
This channel consists of the \ac{LOS} path, which is the path that signals reach the \ac{ue} directly, and/or  \ac{NLOS} paths, which are the paths that signals are reflected by reflecting surfaces or scattered by small objects and then reach the \ac{ue}. The $i$-th multipath can be described by a complex gain $g_{k}^{i}$, a \ac{TOA} $\tau_{k}^{i}$, an \ac{AOA} pair $\boldsymbol{\theta}_{k}^{i}$ in azimuth and elevation, and an \ac{AOD} pair $\boldsymbol{\phi}_{k}^{i}$ in azimuth and elevation, which depend on the hidden geometric relation among the \ac{bs}, \ac{ue} and landmarks, as shown, e.g., in \cite[Appendix A]{ge20205GSLAM}.
Based on the downlink received pilot signals 
at time step $k$, 
the \ac{ue} can execute a channel estimator, for example, \cite{richter2005estimation,Gershman2010,jiang2021high}, and acquire estimates of multipath angles and delays  as measurements, denoted as $\mathcal{Z}_{k}=\{\boldsymbol{z}_{k}^{1},\dots, \boldsymbol{z}_{k}^{\hat{{I}}_{k}} \}$. Due to  clutter and  misdetected landmarks (e.g., due to low path amplitude $|g_{k}^{i}|$ or non-resolvable paths), $\hat{{I}}_{k}$ is usually not equal to real number of multipath $I_{k}$. Affected by  measurement noise, the measurement originating from landmark $\boldsymbol{x}^{i}$ can be modeled as
\begin{align}
    f(\boldsymbol{z}_{k}^{i}|\boldsymbol{x}^{i},\boldsymbol{s}_{k})=\mathcal{N}(\boldsymbol{z}_{k}^{i};\boldsymbol{h}(\boldsymbol{x}^{i},\boldsymbol{s}_{k}),\boldsymbol{R}_k^i),\label{pos_to_channelestimation}
\end{align}
where  $\boldsymbol{R}_k^i$ is the measurement covariance, which can be determined by the Fisher information matrix of channel parameters~\cite{Zohair_5GFIM_TWC2018}, and
$\boldsymbol{h}(\boldsymbol{x}^{i},\boldsymbol{s}_{k})=[\tau_{k}^{i},(\boldsymbol{\theta}_{k}^{i})^{\mathsf{T}},(\boldsymbol{\phi}_{k}^{i})^{\mathsf{T}}]^{\mathsf{T}}$ represents the nonlinear function that transforms the geometric information to the TOA, AOA and AOD.

\section{Methods for Mapping and SLAM} \label{sec:methods}
The objective of  \ac{SLAM} is to determine or approximate the joint posterior density of the \ac{ue} trajectory and landmark states $f(\boldsymbol{s}_{1:k},\mathcal{X}  \lvert \mathcal{Z}_{1:k}, \boldsymbol{u}_{1:k},\boldsymbol{s}_{0})$, given the initial \ac{ue} pose $\boldsymbol{s}_{0}$, measurements $\mathcal{Z}_{1:k}$ and controls $\boldsymbol{u}_{1:k}$ up to and including time instant $k$.  In this section, we first briefly overview the different SLAM methods, including classical methods, GraphSLAM, and AI-based methods. Then, we dive deeper into two methods, namely based on \acp{RFS} and on \ac{BP}, which are natural for dealing with \acp{DA}.

\subsection{Overview of Different Methods}

\runinhead{Classical Methods}
The \ac{SLAM} problem requires solving the joint posterior density of the \ac{ue} and landmarks, conditioned on 
the recorded observations and control inputs up to the current time instant. A common approach 
is to approximate the joint posterior density as a Gaussian distribution and utilizing for example an \ac{EKF} for estimating the posterior  \cite{dissanayake2001}. An alternative solution to SLAM is based on an exact factorization of the posterior into a product of conditional landmark distributions and a distribution over \ac{ue} trajectories, leading to the widely used FastSLAM algorithm \cite{montemerlo2002}. The classical \ac{SLAM} methods solve the problem in two steps. First, the DA is solved and thereafter, Bayesian filtering is used to approximate the posterior assuming the DA does not contain any uncertainty. This two-tiered approach has been demonstrated to work well in practice \cite{dissanayake2001,montemerlo2002}, but it is sensitive to DA uncertainty \cite{neira2001}.

\runinhead{GraphSLAM}
In GraphSLAM, 
the DA problem is usually addressed by the \ac{SLAM} front-end approach \cite{neira2001,nuchter20056d}. Once the DA is known, the GraphSLAM algorithms compute the joint posterior density of the \ac{ue} trajectory and landmarks by transforming the joint posterior density into a graphical network \cite{grisetti2010tutorial}. The nodes in the graphical network represent the \ac{ue} states at different place in time or landmarks, and the edges represent the constrains between two \ac{ue} states, or between the landmark state and the \ac{ue} state, which consists of a probability distribution over the relative transformations between two nodes \cite{thrun2006graph}. These two types of constrains can be obtained by the motion model of the \ac{ue} or the measurement model of the \ac{ue} and landmarks \cite{lu1997globally}. Once the graph is constructed, the \ac{SLAM} problem becomes determining the most likely configuration of these nodes that can maximize the posterior, which can be solved by standard optimization techniques \cite{olson2006fast,grisetti2009nonlinear}. Unlike many other SLAM methods, e.g., EK-SLAM and FastSLAM, which operate online, 
GraphSLAM  performs batch estimation, where the received measurements are  processed batch by batch.

\runinhead{AI-based SLAM}


The combination of deep learning and SLAM, also known as Deep SLAM, is yet to become the default SLAM strategy but has received considerable attention in recent years, especially for vision SLAM (SLAM for vision sensors). The idea is often to maintain a traditional SLAM pipeline but to replace manually designed models and loss functions, which may not be optimal, with optimized neural networks. 
For instance, \cite{Ono2018,yi2016lift} describe methods to use deep learning for key-point detection, and \cite{ranftl2018deep} presents a technique to robustly estimate the fundamental matrix using deep learning. There are also promising attempts at replacing a larger part of the pipeline with deep learning. For instance, \cite{bloesch2018codeslam,teed2021droid} use deep learning to jointly estimate a depth map and a pose for each image, by pairwise comparison of images, and avoid some of the machinery which is common to most vision SLAM solutions. 
Deep learning for mmWave SLAM remains virtually unexplored. However, in spite of the differences between mmWave and vision data, it is still possible that ideas from deep vision SLAM are applicable to mmWave SLAM. As an example, vision SLAM often involves a feature matching problem, where keypoints from different images are associated to each other, which resembles the \ac{DA} problem that we face in mmWave SLAM. SuperGLUE \cite{sarlin2020superglue} demonstrated excellent performance for the feature matching problem using transformer-like architectures \cite{vaswani2017attention}, and related architectures have also been used to handle \acp{DA} in settings that are more closely related to mmWave data \cite{pinto2021next}.

\runinhead{RFS-SLAM}



A \ac{RFS} is a random variable whose possible outcomes are sets with a finite number of unique elements. That means both the number of elements in the set and the elements themselves are random. Therefore, unlike random vectors, where both the number and the order of the elements are pre-fixed,  \acp{RFS} are invariant to the order of elements and  can easily add or remove elements  \cite{mahler2003multitarget,mullane2011random,mahler2014advances}. These merits make the \ac{RFS} particularly attractive to be used to model the unknown environment and the detected measurements in \ac{SLAM} problems, since uncertainty in both the number of landmarks/measurements as well as their individual states can be inherently modeled. In \ac{RFS}-based \ac{SLAM} methods,  different \acp{RFS}, for example, the \ac{LMB} \ac{RFS} in \cite{deusch2015labeled,deusch2016random}, the \ac{GLMB} \ac{RFS} in \cite{moratuwage2018delta,moratuwage2019delta},  the \ac{PPP} \ac{RFS} in \cite{kim20205g,mullane2011random} and the \ac{PMBM} \ac{RFS} in \cite{ge20205GSLAM,ge2021computationally}, are utilized to model the unknown environment. 
To solve the \ac{SLAM} problem, the  joint posterior density of the \ac{ue} and landmarks can be approximated by using particles  \cite{ge20205GSLAM,ge2020exploiting}, or sigma-points  \cite{kim2020a,kim2020b}, or relying on  \ac{EKF}  \cite{ge2021computationally,EKPHD2021Ossi}, or \ac{IPLF}  \cite{ge2021iterated}.

\runinhead{\ac{BP}-based SLAM with Factor Graphs}

    The above 
    \ac{SLAM} problem can be efficiently solved  by computing the \ac{MPD} of the hidden random variables, including the \ac{DA}.
    The \ac{MPD}s for \ac{SLAM} are approximately determined by the \ac{BP} (i.e., message passing) with the sum-product algorithm~\cite{Frank_SPA_TIT2001}, known as BP-SLAM~\cite{Erik_BPSLAM_TWC2019,Erik_AOABPSLAM_ICC2019}.
    Using the random vector instead of RFS, a landmark state is represented, developed from \ac{BP}-based \ac{MTT}~\cite{meyer2018message}.
    To handle the unknown number of landmarks, the augmented random vector (consisting of the location and existence variable) is modeled.
    In \ac{BP}-\ac{SLAM}, the correlation between \ac{ue} state and landmark state cannot be tracked since the hidden variables are marginalized by \ac{BP}.
    The landmarks are divided into two parts: undetected landmarks and detected landmarks, similarly to \ac{PMBM}-\ac{SLAM}.
    However, the Poisson part is partially adopted for modeling undetected landmarks, implicitly requiring additional process outside the \ac{BP} framework.
    Nevertheless, \ac{BP}-\ac{SLAM} is powerful due to its generality to the complex scenario~(e.g., multiple \ac{bs}s).
    BP is also used in some RFS-based SLAM methods, e.g., for computing the marginal probabilities of the \acp{DA}~\cite{williams2015marginal}.
    

\subsection{RFS-SLAM}

The joint posterior \ac{RFS}-\ac{SLAM} density can be decomposed as \cite{mullane2011random} 
\begin{equation}
    f(\boldsymbol{s}_{1:k},\mathcal{X}  \lvert \mathcal{Z}_{1:k}, \boldsymbol{u}_{1:k},\boldsymbol{s}_{0}) = f(\boldsymbol{s}_{1:k}\lvert \mathcal{Z}_{1:k}, \boldsymbol{u}_{1:k},\boldsymbol{s}_{0}) f(\mathcal{X} \lvert \mathcal{Z}_{1:k}, \boldsymbol{s}_{0:k}). \label{eq:jointPDF}
\end{equation}
The recursion for the joint \ac{RFS}-\ac{SLAM} density is equivalent to jointly propagation the posterior density of the \ac{ue} trajectory, $f(\boldsymbol{s}_{1:k}\lvert \mathcal{Z}_{1:k}, \boldsymbol{u}_{1:k},\boldsymbol{s}_{0})$, and the posterior density of the map that is conditioned on the \ac{ue} trajectory, $f(\mathcal{X}  \lvert \mathcal{Z}_{1:k}, \boldsymbol{s}_{0:k})$. Since $\mathcal{X}$ is an \ac{RFS} (i.e., an unordered set of arbitrary cardinality), $f(\mathcal{X}  \lvert \mathcal{Z}_{1:k}, \boldsymbol{s}_{0:k})$ is a so-called \emph{set density}. An example of a set density is a  \ac{PPP} density, given by 
\begin{equation}
    f_{\mathrm{PPP}}(\mathcal{X})=e^{-\int {\lambda}(\boldsymbol{x}')\text{d}\boldsymbol{x}'} \prod_{\boldsymbol{x} \in \mathcal{X}}  {\lambda}(\boldsymbol{x}),\label{PPP}
\end{equation}
where the argument $\mathcal{X}$ can be a set of arbitrary cardinality and  ${\lambda}(\boldsymbol{x})$ denotes the PPP's intensity function. 
Further examples will be provided in Sect.~\ref{PMBMMapRepresentation}.

We consider the point target measurement model, where each landmark can give at most one measurement. Therefore, the RFS likelihood function $g(\mathcal{Z}_{k}|\boldsymbol{s}_{k},\mathcal{X})$ is given by  \cite[eqs.\,(5)--(6)]{garcia2018poisson}
\begin{align}
    & g(\mathcal{Z}_{k}|\boldsymbol{s}_{k},\mathcal{X}) = e^{-\int c(\boldsymbol{z}) \mathrm{d} \boldsymbol{z}}  \sum_{\mathcal{Z}_{k}^c\uplus\mathcal{Z}_{k}^1 \ldots \uplus \mathcal{Z}_{k}^{|\mathcal{X}|}=\mathcal{Z}_{k}}\prod_{\boldsymbol{z} \in \mathcal{Z}^c}c(\boldsymbol{z})\prod_{i=1}^{|\mathcal{X}|}\ell(\mathcal{Z}_{k}^i|\boldsymbol{s}_{k},\boldsymbol{x}^i),\label{likelihood}
\end{align}
where $\mathcal{Z}_{k}^c$ is the clutter measurement set with clutter intensity $c(\boldsymbol{z})$, and 
\begin{align}
    \ell(\mathcal{Z}_{k}^i |\boldsymbol{s}_{k},\boldsymbol{x}^i)=
\begin{cases}
 1-p_{\text{D}}(\boldsymbol{x}^{i},\boldsymbol{s}_{k}) \quad & \mathcal{Z}_{k}^{i}=\emptyset, \\ p_{\text{D}}(\boldsymbol{x}^{i},\boldsymbol{s}_{k})f(\boldsymbol{z}|\boldsymbol{x}^{i},\boldsymbol{s}_{k}) \quad& \mathcal{Z}_{k}^{i}=\{\boldsymbol{z} \}, \\0 \quad & \mathrm{otherwise},
\end{cases}
\end{align}
where $p_{\text{D}}(\boldsymbol{x}^{i},\boldsymbol{s}_{k})\in [0,1]$ is the detection probability, and $f(\boldsymbol{z}|\boldsymbol{x}^{i},\boldsymbol{s}_{k})$ is given by \eqref{pos_to_channelestimation}. 

\runinhead{UE trajectory density}

Given the priors of the UE $f(\boldsymbol{s}_{1:k-1}\lvert \mathcal{Z}_{1:k-1}, \boldsymbol{u}_{1:k-1},\boldsymbol{s}_{0})$ and the map  $f(\mathcal{X}  \lvert \mathcal{Z}_{1:k-1}, \boldsymbol{s}_{0:k-1})$, the likelihood $g(\mathcal{Z}_{k}|\boldsymbol{s}_{k},\mathcal{X})$, and the motion model $f(\boldsymbol{s}_{k}|\boldsymbol{s}_{k-1},\boldsymbol{u}_{k})$, the updated UE density $f(\boldsymbol{s}_{1:k}\lvert \mathcal{Z}_{1:k}, \boldsymbol{u}_{1:k},\boldsymbol{s}_{0})$ can be obtained by
\begin{align}
    &f(\boldsymbol{s}_{1:k}\lvert \mathcal{Z}_{1:k}, \boldsymbol{u}_{1:k},\boldsymbol{s}_{0}) = \frac{ f(\boldsymbol{s}_{1:k}\lvert \mathcal{Z}_{1:k-1}, \boldsymbol{u}_{1:k},\boldsymbol{s}_{0})f(\mathcal{Z}_{k}\lvert\mathcal{Z}_{1:k-1},\boldsymbol{s}_{0:k})}{p(\mathcal{Z}_{k}\lvert\mathcal{Z}_{1:k-1})},
\end{align}
with $f(\boldsymbol{s}_{1:k}\lvert \mathcal{Z}_{1:k-1}, \boldsymbol{u}_{1:k},\boldsymbol{s}_{0})= f(\boldsymbol{s}_{1:k-1}\lvert \mathcal{Z}_{1:k-1}, \boldsymbol{u}_{1:k-1},\boldsymbol{s}_{0})f(\boldsymbol{s}_{k}\lvert  \boldsymbol{u}_{k},\boldsymbol{s}_{k-1})$ denoting the prediction of the \ac{ue} trajectory.

\runinhead{Map density} 
The posterior map  density $f(\mathcal{X}  \lvert \mathcal{Z}_{1:k}, \boldsymbol{s}_{0:k})$ can be obtained by
\begin{align}
    &f(\mathcal{X}  \lvert \mathcal{Z}_{1:k}, \boldsymbol{s}_{0:k}) = \frac{f(\mathcal{X}  \lvert \mathcal{Z}_{1:k-1}, \boldsymbol{s}_{0:k})g(\mathcal{Z}_{k}|\boldsymbol{s}_{k},\mathcal{X})}{f(\mathcal{Z}_{k}\lvert\mathcal{Z}_{1:k-1},\boldsymbol{s}_{0:k})},
\end{align}
where $f(\mathcal{X}  \lvert \mathcal{Z}_{1:k-1}, \boldsymbol{s}_{0:k})=f(\mathcal{X}  \lvert \mathcal{Z}_{1:k-1}, \boldsymbol{s}_{0:k-1})$, as the landmarks are all fixed.

\subsubsection{UE Trajectory Density Representation}
We follow the \ac{RBP} approach \cite[Ch. 7.1]{sarkka2013bayesian}, and approximate the posterior density of the \ac{ue} trajectory using a weighted set of $N$ particles \cite{arulampalam2002}
\begin{equation}
    f(\boldsymbol{s}_{1:k}\lvert \mathcal{Z}_{1:k}, \boldsymbol{u}_{1:k},\boldsymbol{s}_{0}) \approx \sum_{n=1}^N w_k^{(n)} \delta \left(\boldsymbol{s}_k - \boldsymbol{s}_k^{(n)} \right),
\end{equation}
where $\delta(\cdot)$ is the Dirac delta distribution, $\boldsymbol{s}_k^{(n)}$ denotes the $n$-th particle and ${w}_k^{(n)}$ the associated weight. In \ac{RFS}-\ac{SLAM}, the \ac{ue} trajectory is estimated using a \ac{PF} and the posterior \ac{RFS}-\ac{SLAM} density is represented by a weighted set of $N$ particles $\lbrace w_k^{(n)}, \boldsymbol{s}_{0:k}^{(n)}, f(\mathcal{X}  \lvert \mathcal{Z}_{1:k}, \boldsymbol{s}^{(n)}_{0:k})  \rbrace_{n=1}^N.$
In the following, we present two map representations based on \ac{PPP} and \ac{PMBM} \acp{RFS}. Then, we will describe how the particle weights $w_k^{(n)}$ are computed. 


\subsubsection{PHD Map Representation} \label{PHDMapRepresentation}

In \ac{PHD}-\ac{SLAM}, the posterior \ac{RFS} map is approximated by a \ac{PPP} \ac{RFS} following  assumed-density filtering \cite{mullane2011random,minka2001family}
\begin{equation}
    f(\mathcal{X} \vert \mathcal{Z}_{1:k}, \boldsymbol{s}_{0:k}^{(n)}) \approx \frac{\prod_{\boldsymbol{x} \in \mathcal{X}} v_{k \vert k}(\boldsymbol{x} \vert \boldsymbol{s}_{0:k}^{(n)})}{\exp \left( \int v_{k \vert k}(\boldsymbol{x} \vert \boldsymbol{s}_{0:k}^{(n)}) \textrm{d}x \right)},
\end{equation}
where $v_{k \vert k}(\boldsymbol{x} \vert \boldsymbol{s}_{0:k}^{(n)})$ denotes the \ac{PHD} and 
it is equal to the \ac{PPP} intensity.
The \ac{PHD} is parametrized using a \ac{GM} \cite{vo2006}
\begin{equation}\label{eq:phd_gm}
    v^{(n)}_{k \vert k}(\boldsymbol{x} \vert \boldsymbol{s}_{0:k}^{(n)}) = \sum_{i=1}^{M_{k}^{(n)}} \eta_{k}^{(n,i)} \mathcal{N}(\boldsymbol{\mu}_{k}^{(n,i)}, \boldsymbol{\Sigma}_{k}^{(n,i)}),
\end{equation}
where $M_{k}^{(n)}$ is the number of GM components at time $k$ and, $\eta_{k}^{(n,i)}$, $\mu_{k}^{(n,i)}$ and $\Sigma_{k}^{(n,i)}$ are the weight, mean and covariance of landmark $i$ for particle $n$ in corresponding order. In \ac{PHD}-\ac{SLAM}, the trajectory-conditioned map is estimated using a \ac{PHD} filter and the overall \ac{PHD}-\ac{SLAM} density at time $k$ is represented by a set of $N$ particles $  \lbrace w_{k \vert k}^{(n)}, \boldsymbol{s}_{0:k}^{(n)}, v_{k \vert k}^{(n)}(\boldsymbol{x} \vert \boldsymbol{s}_{0:k}^{(n)})  \rbrace_{n=1}^N$. 
It is worth noting that the integral of the PHD over any region gives the expected number of landmarks in that region, and the highest local concentration of expected number of landmarks is captured by local maximas of the PHD \cite{vo2006,mullane2011random}. Next, we summarize the PHD filtering recursion, that is, the prediction and update steps of the filter.

\runinhead{Prediction step} 
 If the PHD at the previous time instant, $v^{(n)}_{k-1 \vert k-1}(\cdot)$, is a GM of the form given in \eqref{eq:phd_gm}, then it follows that the predicted PHD is also a GM given by \cite{vo2006}
\begin{equation}
    v^{(n)}_{k \vert k-1}(\boldsymbol{x} \vert \boldsymbol{s}_{k}^{(n)}) = v^{(n)}_{k-1 \vert k-1}(\boldsymbol{x} \vert \boldsymbol{s}_{k-1}^{(n)}) + v^{(n)}_{\textrm{B},k}(\boldsymbol{x} \vert \boldsymbol{s}_{k}^{(n)}).
\end{equation}
The parameters of $v^{(n)}_{k-1 \vert k-1}(\cdot)$ are unchanged since the landmarks are static, $v^{(n)}_{\textrm{B},k}(\cdot)$ is the birth process with $M_{\textrm{B},k}^{(n)}$ GM components, and the number of components in the predicted PHD is $M_{k \vert k-1}^{(n)} = M_{k-1 \vert k-1}^{(n)} + M_{\textrm{B},k}^{(n)}$. The birth process indicates where and with which intensities new landmarks appear and commonly the birth process is either measurement-driven \cite{kim20205g} or is assumed to be known a priori \cite{vo2006}.

\runinhead{Update step}
 Once measurement set $\mathcal{Z}_k$ is observed at time $k$, the predicted PHD $v^{(n)}_{k \vert k-1}(\cdot)$ can be updated and the posterior map PHD is given by \cite{vo2006,mullane2011random}
 \begin{equation}\label{eq:phd_update}
    v^{(n)}_{k \vert k}(\boldsymbol{x} \vert \boldsymbol{s}_{k}^{(n)}) = \big[1 - p_\textrm{D}(\boldsymbol{x}, \boldsymbol{s}_{k}^{(n)}) \big]v^{(n)}_{k \vert k-1}(\boldsymbol{x} \vert \boldsymbol{s}_{k}^{(n)}) + \sum_{\boldsymbol{z} \in \mathcal{Z}_k} \frac{\boldsymbol{\Lambda}(\boldsymbol{z}, \boldsymbol{x} \vert \boldsymbol{s}_{k}^{(n)}) }{c(\boldsymbol{z}) + \int \boldsymbol{\Lambda}(\boldsymbol{z}, \boldsymbol{x}' \vert \boldsymbol{s}_{k}^{(n)})  \textrm{d}\boldsymbol{x}'},
\end{equation}
where 
\begin{equation}\label{eq:lambda}
  \boldsymbol{\Lambda}(\boldsymbol{z}, \boldsymbol{x} \vert \boldsymbol{s}_{k}^{(n)}) = p_\textrm{D}(\boldsymbol{x}, \boldsymbol{s}_{k}^{(n)}) f(\boldsymbol{z} \vert \boldsymbol{x},\boldsymbol{s}_k^{(n)}) v^{(n)}_{k \vert k-1}(\boldsymbol{x} \vert \boldsymbol{s}_{k}^{(n)}).
\end{equation}
 In \eqref{eq:phd_update}, the first term on the right hand side represents landmarks that are undetected and the latter term is the set of detected landmarks. Since every component of $v^{(n)}_{k \vert k -1}(\cdot)$ is updated by a miss detection and by every measurement,  the number of GM components in the updated PHD $v^{(n)}_{k \vert k}(\cdot)$ is $M_{k \vert k}^{(n)} = M_{k \vert k-1}^{(n)} \times (\vert \mathcal{Z}_k \vert + 1)$. In practice, the weight, mean and covariance of the detected landmarks can be updated using any standard Kalman filtering technique such as the \ac{EKF} and for details how to compute the GM parameters, the readers are referred to \cite{kim20205g, mullane2011random, vo2006}.

\subsubsection{PMBM Map Representation}\label{PMBMMapRepresentation}

\label{sec:Ch3_PMBM}

For notational convenience, we will drop the particle index $n$ in this section. However, please note that all the densities are conditioned on a given particle. 

The \ac{PMBM}-based \ac{SLAM} filter utilizes the \ac{PMBM} \ac{RFS} to represent the environment, conditioned on the \ac{ue} state. In \cite{williams2015marginal,garcia2018poisson}, the \ac{PMBM} density is proven to be a conjugate prior for \eqref{likelihood}. Therefore, the posterior keeps the \ac{PMBM} format if the prior is a \ac{PMBM} density, so that we can directly update the \ac{PMBM} components. 
Unlike the \ac{PHD}-based SLAM filter, where the environment are modeled as a \ac{PPP} \ac{RFS}, the \ac{PMBM} \ac{RFS} considers two types of landmarks, the undetected and detected landmarks. The undetected landmarks are the landmarks that exist but have never been detected until the current time, which are modeled as a \ac{PPP} \ac{RFS} $\mathcal{X}_{\mathrm{U}}$. The  detected landmarks are the landmarks that have been detected at least once until the current time, which are modeled as a \ac{MBM} \ac{RFS} $\mathcal{X}_{\mathrm{D}}$ \cite{williams2015marginal,fatemi2017poisson,garcia2018poisson}. As undetected and detected landmarks are disjoint, a \ac{PMBM} \ac{RFS} $\mathcal{X}$ can be expressed as the union of $\mathcal{X}_{\mathrm{U}}$   and $\mathcal{X}_{\mathrm{D}}$. By applying the \ac{FISST} convolution, the density of the \ac{PMBM} \ac{RFS} $\mathcal{X}$ follows \cite{mahler2014advances} 
\begin{equation}
    f_{\mathrm{PMBM}}(\mathcal{X})=\sum_{\mathcal{X}_{\mathrm{U}}\uplus\mathcal{X}_{\mathrm{D}}=\mathcal{X}}f_{\mathrm{PPP}}(\mathcal{X}_{\mathrm{U}})f_{\mathrm{MBM}}(\mathcal{X}_{\mathrm{D}}),\label{PMBM}
\end{equation}
where $\uplus$ represents the union of mutually disjoint sets, $f_{\mathrm{PMBM}}(\cdot)$ is the \ac{PMBM} density, $f_{\mathrm{PPP}}(\cdot)$ is the \ac{PPP} density and $f_{\mathrm{MBM}}(\cdot)$ is the \ac{MBM} density. 
The PPP density 
has intensity 
${\lambda}(\boldsymbol{x})=\mu {f}_{\mathrm{u}}(\boldsymbol{x})$. It represents that the cardinality of $\mathcal{X}_{\mathrm{U}}$ (the number of undetected landmarks) follows a Poisson distributed with given mean $\mu$, and all undetected landmarks are independently and identically distributed with a given spatial density ${f}_{\mathrm{u}}(\boldsymbol{x})$.
The \ac{MBM} density considers multiple global hypotheses (more on this soon) with multiplicative weights such that
\begin{equation}
    f_{\mathrm{MBM}}(\mathcal{X}_{\mathrm{D}})= \sum_{j\in\mathbb{I}}\omega^{j}\sum_{\mathcal{X}^{1}\uplus \dots \uplus
    \mathcal{X}^{n}=\mathcal{X}_{\mathrm{D}}}\prod_{i=1}^{n}f^{j,i}_{\mathrm{B}}(\mathcal{X}^{i}),\label{MBM}
\end{equation}
where $j$ is an index in the set of global hypotheses $\mathbb{I}$ \cite{williams2015marginal}; $\omega^{j}$ is the weight for global hypothesis $j$, satisfying $\sum_{j\in\mathbb{I}}\omega^{j}=1$; 
$n$ is the number of potentially detected landmarks; and $f_{\mathrm{B}}^{j,i}(\cdot)$ is the Bernoulli density of landmark $i$ under global hypothesis $j$, following
\begin{equation}
f^{j,i}_{\mathrm{B}}(\mathcal{X}^{i})=
\begin{cases}
1-r^{j,i} \quad& \mathcal{X}^{i}=\emptyset, \\ r^{j,i}f^{j,i}(\boldsymbol{x}) \quad & \mathcal{X}^{i}=\{\boldsymbol{x}\}, \\ 0 \quad & \mathrm{otherwise},
\end{cases}
\end{equation} 
where $r^{j,i} \in [0,1]$ is the existence probability, representing how likely the landmark exists, and $f^{j,i}(\cdot)$ is its state density if exists, which is assumed to be a Gaussian. A higher $r^{j,i}$ represents the corresponding landmark is more likely to exist, and a lower $r^{j,i}$ represents the corresponding landmark is more likely to not exist. Please note that if $\mathcal{X}^{i}$ is empty (the corresponding landmark does not exist) under the $j$-th global hypothesis, $r^{j,i}$ is 0, resulting the $f^{j,i}_{\mathrm{B}}(\mathcal{X}^{i})=1$.


By plugging \eqref{PPP} and \eqref{MBM} into \eqref{PMBM}, we then can rewrite \eqref{PMBM} as 
\begin{equation}
    f_{\mathrm{PMBM}}(\mathcal{X})=\sum_{\mathcal{X}_{\mathrm{U}}\uplus\mathcal{X}_{\mathrm{D}}=\mathcal{X}}e^{-\int {\lambda}(\boldsymbol{x}')d\boldsymbol{x}'} \prod_{\boldsymbol{x} \in \mathcal{X}_{\mathrm{U}}}  {\lambda}(\boldsymbol{x})\sum_{j\in\mathbb{I}}\omega^{j}\sum_{\mathcal{X}^{1}\uplus \dots \uplus \mathcal{X}^{n}=\mathcal{X}_{\mathrm{D}}}\prod_{i=1}^n
    f^{j,i}_{\mathrm{B}}(\mathcal{X}^{i}).\label{PMBM_details}
\end{equation}
Then, \eqref{PMBM_details} can be parameterized by
$\lambda(\boldsymbol{x})$ and  $\{\omega^{j},\{r^{j,i},f^{j,i}(\boldsymbol{x})\}_{i\in \mathbb{I}^{j}}\}_{j\in \mathbb{I}}$, where $\mathbb{I}^{j}$ is the index set of all considered landmarks under global hypotheses $j$.

\runinhead{Global and local hypotheses}
Each detected landmark has been associated to at least one measurement. However, that measurements may have been a false alarm, so the landmark  should be interpreted as a potentially existing landmark with a certain existence probability, which motivates the use of a Bernoulli density.

Given the measurements $\mathcal{Z}$, there are two cases for each previously detected landmark: (i) the landmark can be associated to a measurement; (ii) the landmark is not associated to any measurement (it is misdetected). Similarly, given the previously detected landmarks, there are two cases for each measurement: (a) it is associated to a previously detected landmark; (b) it is not associated to any of the previously detected landmarks, but comes from clutter or a newly detected landmark. 

\begin{figure}
    \centering
    \includegraphics[width=1\textwidth]{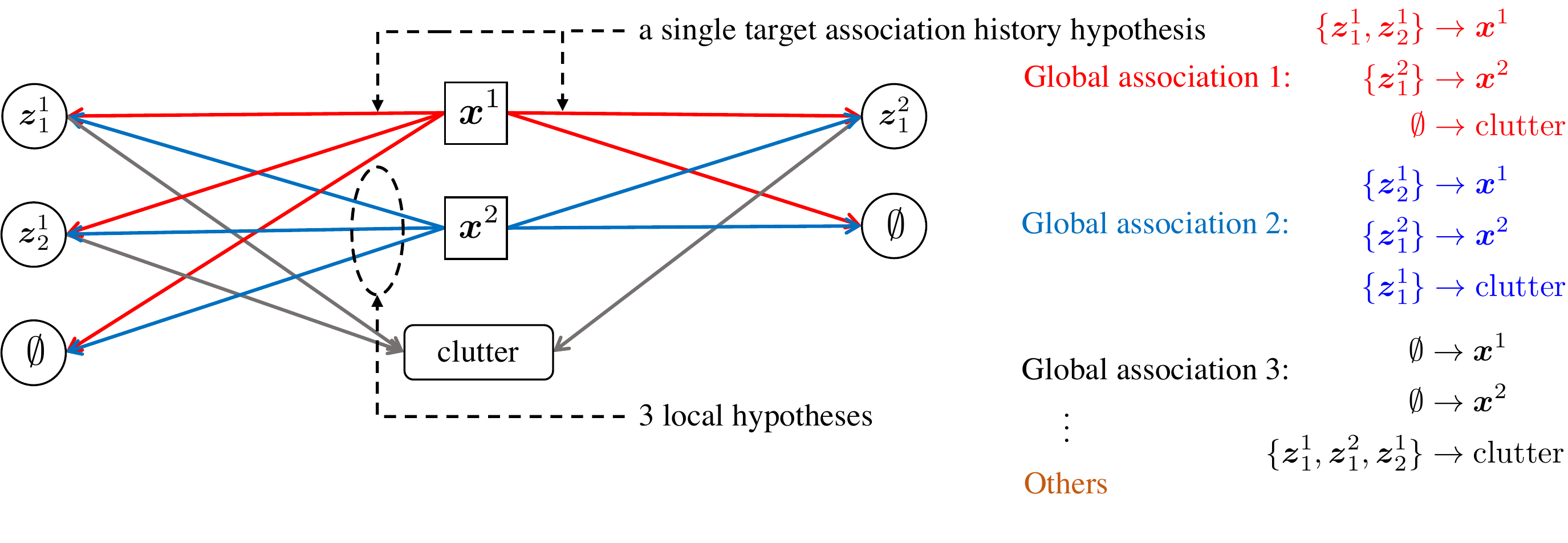}
    \caption{Example of the global hypotheses of associating measurement sets at two time step $\mathcal{Z}_{1}=\{\boldsymbol{z}_{1}^{1},\boldsymbol{z}_{1}^{2}\}$ and $\mathcal{Z}_{2}=\{\boldsymbol{z}_{2}^{1}\}$ to two landmarks $\boldsymbol{x}^{1}$ and $\boldsymbol{x}^{2}$. Examples of the local hypotheses and the single target association history hypothesis are also shown. Each single measurement can be associated to a landmark or a clutter, and each single landmark can be detected with a measurement each time or misdetected (associated with $\emptyset$ in the figure).}
    \label{fig:DA}\vspace{-4mm}
\end{figure}

In  other words, a single measurement can be associated either to a previously detected landmark, or a newly detected landmark or  clutter, which are the local hypotheses. 
The history of the local hypotheses of a potentially detected landmark, which is referred to  a `single target association history hypothesis', incorporates information on when it was first detected and by which measurement, when it is misdetected, and when it is detected again with which measurement. 
Single target association history hypotheses are generated per landmark, but may be globally inconsistent among landmarks (e.g., a measurement may be assigned to two landmarks). 
A global hypotheses, which is also known as a valid \ac{DA}, contains one single target association history hypothesis for each potentially detected landmark,  with the constraint that each measurement is contained in only one of the single target association history hypotheses \cite{williams2015marginal}. Fig.~\ref{fig:DA} visualizes the local hypotheses, a single target association history hypothesis and the global hypothesis of an exemplary association problem. The summation in \eqref{MBM} is over the global hypotheses. 

\runinhead{Prediction step}
Assume that the \ac{PMBM} at time $k-1$ is the form of \eqref{PMBM_details} with parameters ${\lambda}_{k-1}(\boldsymbol{x})$, and $\{\omega^{j}_{k-1},\{r^{j,i}_{k-1},f^{j,i}_{k-1}(\boldsymbol{x})\}_{i\in \mathbb{I}_{k-1}^{j}}\}_{j\in \mathbb{I}_{k-1}}$. 
As all landmarks are fixed, there is no prediction step, and the posterior map \ac{PMBM} can be directly update by $\mathcal{Z}_{k}$, which is index by $p \in \{1,2,\cdots, |\mathcal{Z}_{k}|\}$. Please note that the \ac{PHD} filter considers the newly detections in the prediction step, while the \ac{PMBM} filter consider them in the update step.

\runinhead{Update step}
In the \ac{PMBM} update step, measurements are used to correct the landmark states. Four different cases are considered \cite{garcia2018poisson}, based on the local hypotheses (i)--(ii) and (a)--(b) discussed above:
\begin{itemize}
\item[a)] \emph{Undetected landmarks that remain undetected:} As the previously undetected landmarks can be still undetected with  possibility  $1 - p_\textrm{D}(\boldsymbol{x})$, the updated intensity for remaining undetected landmarks is given by
\begin{equation}
    {\lambda}_{k}(\boldsymbol{x}) = (1 - p_\textrm{D}(\boldsymbol{x})) {\lambda}_{k-1}(\boldsymbol{x}).
\end{equation}
\item[b)] \emph{Previously undetected landmark is detected for the first time with the measurement $\boldsymbol{z}_{k}^{p}$:} A new Bernoulli will be created by the measurement $\boldsymbol{z}_{k}^{p}$ with state density $f_{k}^{p}(\boldsymbol{x}\vert \boldsymbol{z}_{k}^{p})$,  existence probability $r_{k}^{p}$ and association weight $l_{k}^{p}$ such as:
\begin{align}
    &r_{k}^{p}=\frac{\int p_{\mathrm{D}}(\boldsymbol{x})f(\boldsymbol{z}_{k}^{p}|\boldsymbol{x}) {\lambda}_{k-1}(\boldsymbol{x}) \text{d}\boldsymbol{x}}{c(\boldsymbol{z})+\int p_{\mathrm{D}}(\boldsymbol{x})f(\boldsymbol{z}_{k}^{p}|\boldsymbol{x}) {\lambda}_{k-1}(\boldsymbol{x}) \text{d}\boldsymbol{x}},\\
    &f_{k}^{p}(\boldsymbol{x}\vert \boldsymbol{z}_{k}^{p})=\frac{p_{\mathrm{D}}(\boldsymbol{x})f(\boldsymbol{z}_{k}^{p}|\boldsymbol{x}) {\lambda}_{k-1}(\boldsymbol{x})}{\int p_{\mathrm{D}}(\boldsymbol{x})f(\boldsymbol{z}_{k}^{p}|\boldsymbol{x}) {\lambda}_{k-1}(\boldsymbol{x}) \text{d}\boldsymbol{x}},\\
    &l_{k}^{p}=c(\boldsymbol{z})+\int p_{\mathrm{D}}(\boldsymbol{x})f(\boldsymbol{z}_{k}^{p}|\boldsymbol{x}) {\lambda}_{k-1}(\boldsymbol{x}) \text{d}\boldsymbol{x},\label{first_weight}
\end{align}
where $r_{k}^{p}\le 1$, which is due to the measurement can come from a real landmark (the landmark exists), or it can be a clutter (the landmark non-exists). Note that $j$ and $i$ do not appear, since the object was undetected before and it is not contained in any previous global hypothesis.

\item[c)] \emph{Previously detected landmark that is misdetected:} A landmark not being associated to any measurements may be due to imperfect detection performance at the sensor or because the landmark in fact does not exist. Then, the updated Bernoulli with association weight $l_{k}^{l,i,0}$ will have a reduced existence and the same state density:
\begin{align}
    &r_{k}^{j,i,0}=\frac{r_{k-1}^{j,i}\int(1-p_{\mathrm{D}}(\boldsymbol{x}))f_{k-1}^{j,i}(\boldsymbol{x})\text{d}\boldsymbol{x}}{1-r_{k-1}^{j,i}+r_{k-1}^{j,i}\int(1-p_{\mathrm{D}}(\boldsymbol{x}))f_{k-1}^{j,i}(\boldsymbol{x})\text{d}\boldsymbol{x}},\label{mis_exist}\\
    &f_{k}^{j,i,0}(\boldsymbol{x})=f_{k-1}^{j,i}(\boldsymbol{x}),\\
    &l_{k|k}^{j,i,0}=1-r_{k-1}^{j,i}+r_{k-1}^{j,i}\int(1-p_{\mathrm{D}}(\boldsymbol{x}))f_{k-1}^{j,i}(\boldsymbol{x})\text{d}\boldsymbol{x}\label{misdetect_weight},
\end{align}
where the index $0$ represents for no measurement being associated to the landmark. The numerator of \eqref{mis_exist} denotes the probability of landmark $\boldsymbol{x}_{k-1}^{j,i}$ exists, but not detected, and  $(1-r_{k-1}^{j,i})$ in the denominator of \eqref{mis_exist} denotes the probability that landmark $\boldsymbol{x}_{k-1}^{j,i}$ does not exist. 

\item[d)] \emph{Previously detected landmark that is detected again:} If a previously detected landmark is detected again with any measurement $\boldsymbol{z}_{k}^{p}$, 
then the existence and the state density of the corresponding updated Bernoulli and its association weight are:
\begin{align}
    &r_{k}^{j,i,p}=1,\\
    &f_{k}^{j,i,p}(\boldsymbol{x}\vert \boldsymbol{z}_{k}^{p})= \frac{p_{\mathrm{D}}(\boldsymbol{x}) f(\boldsymbol{z}_{k}^{p}|\boldsymbol{x})f_{k-1}^{j,i}(\boldsymbol{x})}{\int p_{\mathrm{D}}(\boldsymbol{x})f(\boldsymbol{z}_{k}^{p}|\boldsymbol{x})f_{k-1}^{j,i}(\boldsymbol{x})\text{d}\boldsymbol{x}},\\
    &l_{k}^{j,i,p}= r_{k-1}^{j,i}\int p_{\mathrm{D}}(\boldsymbol{x})f(\boldsymbol{z}_{k}^{p}|\boldsymbol{x})f_{k-1}^{j,i}(\boldsymbol{x})\text{d}\boldsymbol{x}.\label{detect_again_weight}
\end{align}
\end{itemize}

At this point, we have calculated all possible local associations of each landmark and each measurement, as well as the local association weights. However, we still need to form the global hypotheses. To form new global hypotheses, under each previous global hypothesis, we need to go through all possible \acp{DA}. Each valid possibility will give rise a new global hypothesis. This will make the number of global hypotheses increase combinatorially for 
each particle, which brings high computational cost
\cite[Appendix B]{ge20205GSLAM}. 
To avoid rapidly increasing global hypotheses, we can approximate this update by using Murty’s algorithm \cite{murty1968letter}, which keeps $\gamma\ge 1$ best global hypotheses with highest likelihoods for each previous global hypothesis. 

For previous global hypothesis $j$, measurements should be assigned to existing Bernoullis or newly created Bernoullis, with the constrains that one measurement can only to associated to one Bernoulli and each Bernoulli can be associated to at most one measurement. Therefore, we can construct a cost matrix using association weights $l_{k}^{p}$, $l_{k}^{j,i,0}$ and $l_{k}^{j,i,p}$ calculated in \eqref{first_weight}, \eqref{misdetect_weight}, \eqref{detect_again_weight} as \cite[eq. (47)]{garcia2018poisson}
\begin{align}
&\mathbf{L}^{j}_{k} =  -\ln \left[
\begin{matrix}
\tilde{l}_{k}^{j,1,1} & \ldots & \tilde{l}_{k}^{j,|\mathbb{I}^{j}_{k-1}|,1} \\ 
\vdots &  \ddots & \vdots \\
\tilde{l}_{k}^{j,1,|\mathcal{Z}_{k}|}  & \ldots & \tilde{l}_{k}^{j,|\mathbb{I}^{j}_{k-1}|,|\mathcal{Z}_{k}|}
\end{matrix}
\left|
\,
\begin{matrix}
l^{1}_{k}  & \ldots & 0 \\ 
\vdots &  \ddots & \vdots \\
0 &  \ldots & l^{|\mathcal{Z}_{k}|}_{k}
\end{matrix}
\right.
\right],
\end{align}
where $\tilde{l}^{j,i,p}_{k}=l^{j,i,p}_{k}/l^{j,i,0}_{k}$. The left $|\mathcal{Z}_{k}| \times |\mathbb{I}^{j}_{k-1}|$ sub-matrix in $\mathbf{L}^{j}_{k}$ corresponds to previously detected landmarks, the right $|\mathcal{Z}_{k}| \times |\mathcal{Z}_{k}|$ diagonal sub-matrix corresponds to newly detected landmarks, and the off-diagonal elements of the right sub-matrix are $-\infty$. The $\gamma$-best \acp{DA} with weights 
can be selected out by solving the assignment problem 
\begin{align}\label{optimization_problem}
\text{minimize} \quad  & \text{tr} \left(\mathbf{A}^{\mathsf{T}} \mathbf{L}^{j}_{k} \right) \\
\text{s.t.} \quad  & [\mathbf{A} ]_{\alpha,\beta} \in 
\{ 0,  1 \} \quad \forall \; \alpha,\beta \nonumber \\ 
& \sum\nolimits_{\beta=1}^{|\mathbb{I}^{j}_{k-1}| + |\mathcal{Z}_{k}|} [ \mathbf{A} ]_{\alpha,\beta} = 1, \quad \forall \; \alpha \nonumber \\ 
& \sum\nolimits_{\alpha=1}^{|\mathcal{Z}_{k}|} [\mathbf{A} ]_{\alpha,\beta} \in  \{ 0,  1 \}, \quad \forall \; \beta \nonumber
\end{align}
using the Murty's algorithm \cite{murty1968letter}, where  $\mathbf{A}\in \{0,1\}^{|\mathcal{Z}_{k}| \times (|\mathcal{Z}_{k}|+|\mathbb{I}^{j}_{k-1}|)}$ is the assignment matrix. The solutions are denoted by $\mathbf{A}^{j,h}$, where $h$ is an index
in the index set of new \acp{DA} under global hypothesis $j$, denoted as $\mathbb{H}^{j}_{k}$ with $|\mathbb{H}^{j}_{k}|\leq \gamma$. Each new \ac{DA} has a weight $\omega^{j,h}_{k}$, given by 
\begin{align}
    \omega^{j,h}_{k} \propto \omega^{j}_{k-1} e^{-\text{tr} \left((\mathbf{A}^{j,h})^{\mathsf{T}}\mathbf{L}^{j}_{k}\right)}
\end{align}
with $\sum_{j\in \mathbb{I}_{k-1}}\sum_{h \in \mathbb{H}^{j}_{k}} \omega^{j,h}_{k}=1$. The index set of landmarks under the $j,h$-th ``new \ac{DA}'' is denoted as $\mathbb{I}_{k}^{j,h}$, with $|\mathbb{I}_{k}^{j,h}|\leq |\mathbb{I}_{k-1}^{j}|+|\mathcal{Z}_{k}|$, as some new birth components may not exist under some \acp{DA}.

After update,  the map follows the PMBM format, with PPP intensity as ${\lambda}_{k}(\boldsymbol{x})$ and  MBM components as $\{\{\omega^{j,h}_{k},\{r^{j,h,i}_{k},f^{j,h,i}_{k}(\boldsymbol{x})\}_{i\in \mathbb{I}_{k}^{j,h}}\}_{h\in \mathbb{H}_{k}^{j}}\}_{j\in \mathbb{I}_{k-1}}$. For the \ac{MBM}, all \acp{DA} can be represented by only using one index. Hence, we reorder all \acp{DA} using index set $ \mathbb{I}_{k}=\{1,\cdots,\sum_{j\in \mathbb{I}_{k}}|\mathbb{H}^{j}_{k}|\}$. Then, MBM components can also be written as $\{\omega^{j}_{k},\{r^{j,i}_{k},f^{j,i}_{k}(\boldsymbol{x})\}_{i\in \mathbb{I}_{k}^{j}}\}_{j\in \mathbb{I}_{k}}$. 

\subsubsection{UE Trajectory Weight Computation}

The posterior of the UE trajectory can be recursively approximated  using the \ac{SIR} \ac{PF} for which the weight update is given by \cite{arulampalam2002}
\begin{equation}\label{eq:pf_particle_weight}
    w_k^{(n)} = w_{k-1}^{(n)} \frac{f(\mathcal{Z}_k \vert \mathcal{Z}_{1:k-1}, \boldsymbol{s}_{0:k}^{(n)}) f(\boldsymbol{s}_k^{(n)} \vert \boldsymbol{s}_{k-1}^{(n)}, \boldsymbol{u}_k)}{q(\boldsymbol{s}^{(n)}_k \vert \boldsymbol{s}_{0:k-1}^{(n)}, \mathcal{Z}_{1:k}, \boldsymbol{u}_{1:k})}.
\end{equation}
A typical choice for the importance density $q(\cdot)$ is the motion model in \eqref{transition_density} \cite{kim20205g,mullane2011random,ge20205GSLAM}, simplifying \eqref{eq:pf_particle_weight} to 
\begin{equation}\label{eq:pf_particle_weight_simplified}
   w_k^{(n)} = w_{k-1}^{(n)} f(\mathcal{Z}_k \vert \mathcal{Z}_{1:k-1}, \boldsymbol{s}_{0:k}^{(n)}). 
\end{equation}
After updating the weight for each particle, the weights are normalized, $w_k^{(n)} = w_k^{(n)} / \sum_{n=1}^N w_k^{(n)}$, and resampling is performed every time step. 

The computation of $f(\mathcal{Z}_k \vert \mathcal{Z}_{1:k-1}, \boldsymbol{s}_{0:k}^{(n)})$ is different for the \ac{PHD} and \ac{PMBM} filters. With a \ac{PPP} prior and a point object measurement model the term for the \ac{PHD} filter is given by \cite{kim20205g}
\begin{equation}\label{eq:measurement_weight}
    f(\mathcal{Z}_k \vert \mathcal{Z}_{1:k-1}, \boldsymbol{s}_{0:k}^{(n)}) = \prod_{\boldsymbol{z} \in \mathcal{Z}_k} \left( c(\boldsymbol{z}) + \int \boldsymbol{\Lambda}(\boldsymbol{z}, \boldsymbol{x} \vert \boldsymbol{s}_{k}^{(n)})  \textrm{d}\boldsymbol{x} \right),
\end{equation}
and it can be evaluated during the PHD update step. Correspondingly, for the \ac{PMBM} filter, the term can be acquired by
\cite[Sect.~III.D]{garcia2018poisson}
\begin{align}
    f(\mathcal{Z}_k \vert \mathcal{Z}_{1:k-1}, \boldsymbol{s}_{0:k}^{(n)}) & \approx  e^{-\int {\lambda}^{(n)}_{k-1}(\boldsymbol{x}')\text{d}\boldsymbol{x}'-\int c(\boldsymbol{z}')\text{d}\boldsymbol{z}'}\label{normalized_item} \\&\times \sum_{j\in \mathbb{I}^{(n)}_{k-1}} \omega^{(n),j}_{k-1} \prod_{i \in \mathbb{I}_{k-1}^{(n),j}} l^{(n),j,i,0}_{k} \sum_{h \in \mathbb{H}^{(n),j}_{k}} e^{-\text{tr} \left((\mathbf{A}^{(n),j,h})^{\mathsf{T}}\mathbf{L}^{(n),j}_{k}\right)}. \nonumber
\end{align}
In \eqref{normalized_item}, the approximation is because not all possible \acp{DA} are considered, and $e^{-\int {\lambda}^{(n)}_{k-1}(\boldsymbol{x}')\text{d}\boldsymbol{x}'}$ and $e^{-\int c(\boldsymbol{z}')\text{d}\boldsymbol{z}'}$ are the normalization items of \ac{PPP} densities for previously undetected landmarks and clutter, respectively.

\subsubsection{Complexity Reduction}

It is worth noting that the above mentioned two RFS-SLAM filters have high complexity, due to the exponential growth in the number of particles with the dimension of the \ac{ue} state, in the number of GM components in the updated PHD with time, and in the number of global hypotheses in the updated \ac{PMBM} with time (if all global hypotheses are kept). Therefore, the complexity should be reduced in practical implementation. 

For example, instead of following the \ac{RBP} approach, we can approximate the joint posterior by using sigma-points \cite{kim2020a,kim2020b}, or relying on linearization \cite{ge2021computationally,EKPHD2021Ossi,ge2021iterated}. In \ac{PHD}-based SLAM filters, GM components with weight lower than a set threshold can be pruned and similar GM components can be merged \cite{vo2006}. To reduce the complexity caused by the \ac{PMBM} density, pruning those global hypotheses with low weights and/or keeping a certain number of global hypotheses can be used for reducing the number of global hypotheses. If such methods are applied, weights should be re-normalized. Bernoullis with low existence probabilities can be pruned or recycled, and the mixture intensity for the \ac{PPP} part can also be pruned and/or merged \cite{williams2015marginal,garcia2018poisson,williams2012hybrid}. Additionally, to further reduce the complexity, a \ac{PMB} can be approximated from the resulting \ac{PMBM} by applying algorithms, for example, the \ac{tomb}, the \ac{momb} \cite{williams2015marginal}, and the Kullback-Leibler
divergence minimization  algorithms \cite{williams2014efficient}, which reduces the number of hypotheses to one after each update.



\subsection{\ac{BP}-based SLAM with Factor Graphs}\label{sec:BP-SLAM}


The principle of BP-SLAM is to consider only vector random variables and not set random variables, and condition the  joint density \eqref{eq:jointPDF} (after casting in vector form) on the global association. Then, by marginalization, the approximate posterior of the UE state and landmark states can be recovered with low computational complexity. 

\subsubsection{Basics of Factor Graph and \ac{BP}}

\begin{figure}[t!]
    \centering
    \includegraphics[width=0.7\textwidth]{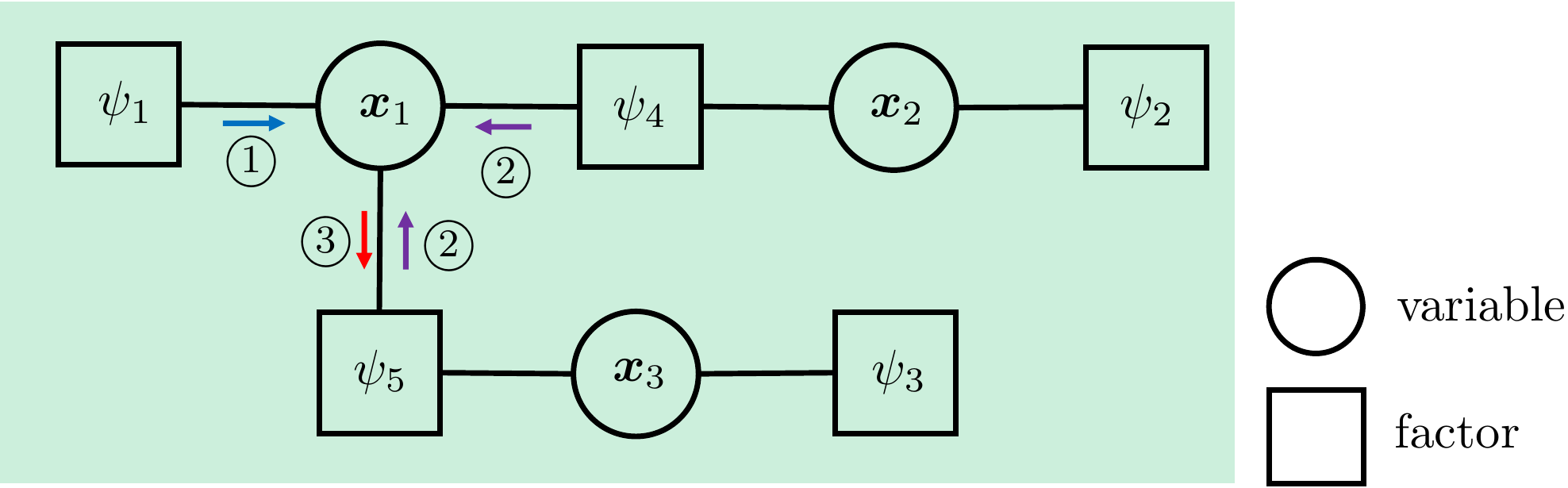}
    \caption{Factor graph representation of  $f(\boldsymbol{x}|\boldsymbol{z})\propto \psi_1(\boldsymbol{x}_1)\psi_2(\boldsymbol{x}_2)\psi_3(\boldsymbol{x}_3)\psi_4(\boldsymbol{x}_1,\boldsymbol{x}_2)\psi_5(\boldsymbol{x}_1,\boldsymbol{x}_3)$. The variables and factors are respectively represented by the circles and squares.}
    \label{fig:Ch3_FGbasic}\vspace{-4mm}
\end{figure}

    \ac{BP}, also known as the sum-product algorithm, is used in marginalizing the posterior density $f(\boldsymbol{x}|\boldsymbol{z})$ that can be factorized as the product of factors, expressed as
\begin{align}
    f(\boldsymbol{x}|\boldsymbol{z}) \propto \prod_{i=1}^I 
    \psi_i(\boldsymbol{x}^{(i)}).
\end{align}
    For the sake of argument, let us consider
    the posterior density that can be factorized as $f(\boldsymbol{x}|\boldsymbol{z})\propto \psi_1(\boldsymbol{x}_1)\psi_2(\boldsymbol{x}_2)\psi_3(\boldsymbol{x}_3)\psi_4(\boldsymbol{x}_1,\boldsymbol{x}_2)\psi_5(\boldsymbol{x}_1,\boldsymbol{x}_3)$, where
    $I=5$, $\boldsymbol{x}^{(1)} = \boldsymbol{x}_1$, $\boldsymbol{x}^{(2)} = \boldsymbol{x}_2$, $\boldsymbol{x}^{(3)} = \boldsymbol{x}_3$, $\boldsymbol{x}^{(4)} = [\boldsymbol{x}_1^\top,\boldsymbol{x}_2^\top]^\top$, $\boldsymbol{x}^{(5)} = [\boldsymbol{x}_1^\top,\boldsymbol{x}_3^\top]^\top$. 
    The factor graph of the factorization is represented by the variable $\boldsymbol{x}_j$ and factor $\psi_i$, depicted in Fig.~\ref{fig:Ch3_FGbasic}.
    By running \ac{BP} on the factor graph, we can compute two types of messages~(from the factor to variable; and from the variable to factor) and belief at each variable. The message from factor $i$ to variable $j$ is denoted by $\mathsf{m}_{\psi_i \rightarrow j}(\boldsymbol{x}_j)$, message from variable $j$ to factor $i$ is denoted by $\mathsf{m}_{j \rightarrow \psi_i}(\boldsymbol{x}_j)$, belief at  variable $j$ is denoted by $\mathsf{b}(\boldsymbol{x}_j)$.
    Examples for messages near variable $1$ and belief at variable $1$~(see, Fig.~\ref{fig:Ch3_FGbasic}) are provided as follows.
    Message \large\protect\textcircled{1}\normalsize~is given by 
    $\mathsf{m}_{\psi_1 \rightarrow 1}(\boldsymbol{x}_1) = \psi_1(\boldsymbol{x}_1)$, 
    two message \large\protect\textcircled{2}\normalsize~are respectively given by $\mathsf{m}_{\psi_4 \rightarrow 1}(\boldsymbol{x}_1) = \int \mathsf{m}_{2 \rightarrow \psi_2 }(\boldsymbol{x}_2) \psi_4(\boldsymbol{x}_1,\boldsymbol{x}_2) \mathrm{d} \boldsymbol{x}_2$ and $\mathsf{m}_{\psi_5 \rightarrow 1}(\boldsymbol{x}_1) = \int \mathsf{m}_{\psi_3 \rightarrow 3}(\boldsymbol{x}_3) \psi_5(\boldsymbol{x}_1,\boldsymbol{x}_3) \mathrm{d} \boldsymbol{x}_3$, and message \large\protect\textcircled{3}\normalsize~is given by $\mathsf{m}_{1 \rightarrow \psi_5}(\boldsymbol{x}_1)= \mathsf{m}_{\psi_1 \rightarrow 1}(\boldsymbol{x}_1)\mathsf{m}_{\psi_4 \rightarrow 1}(\boldsymbol{x}_1)$.
    The belief $\mathsf{b}(\boldsymbol{x}_1)$ indicates update of \ac{BP} at variable $1$, given by $\mathsf{b}(\boldsymbol{x}_1) \propto \mathsf{m}_{\psi_1 \rightarrow 1}(\boldsymbol{x}_1)\mathsf{m}_{\psi_4 \rightarrow 1}(\boldsymbol{x}_1)\mathsf{m}_{\psi_5 \rightarrow 1}(\boldsymbol{x}_1)$ \cite{Frank_SPA_TIT2001}.
    
\subsubsection{Factorization and Factor Graph of Joint SLAM Density}
    Using the vector representation instead of the set representation, a landmark variable $\boldsymbol{y}=[\boldsymbol{x}^\top, \epsilon]^\top$ is introduced, where $\boldsymbol{x}$ denotes the landmark state, and $\epsilon\in\{0,1\}$ denotes the existence variable.
    Then, the \ac{PDF} of a landmark state is denoted by $f(\boldsymbol{x}, \epsilon)$.
    The previously and newly detected landmark are respectively marked by $\tilde{\cdot}$ and $\breve{\cdot}$: for instance, $\tilde{\boldsymbol{x}}$ and $\breve{\boldsymbol{x}}$.
    
    We denote by $I_{k-1}$ and $J_k$, respectively, the number of previously detected landmarks and the number of measurements. 
    We introduce two association variables $\boldsymbol{c}_k = [c_k^1,...,c_k^{I_{k-1}}]$, $\boldsymbol{d}_k = [d_k^1,...,d_k^{J_k}]$.
    Here, $c_k^i \in \{0,1,\ldots,J_k\}$ denotes the association of landmark $i$ with measurement $j$, and $d_k^j \in \{0,1,\ldots,I_{k-1}\}$ denotes the association of measurement $j$ with landmark $i$.
    We also introduce the association factor $\psi(c_k^i,d_k^j)$, modeled to be 0 if the association is not valid~(i.e., $c_k^i = j$, $d_k^j \neq i$ or $c_k^i \neq j$, $d_k^j = i$), and 1 otherwise~\cite{williams2015marginal}.
    
    With the above association variable and vector representation, we introduce two variants likelihood functions from~\cite{garcia2018poisson}: likelihood function for the undetected landmarks, denoted by $\tilde{l}(\boldsymbol{z}_k^j | \boldsymbol{s}_k, \breve{\boldsymbol{x}}_k,\breve{\epsilon}_k, d_k^j)$, given by
\begin{align}
    \tilde{l}(\boldsymbol{z}_k^j | \boldsymbol{s}_k, \breve{\boldsymbol{x}}_k,\breve{\epsilon}_k, d_k^j) 
    &=
    \begin{cases}
        p_\mathrm{D}(\breve{\boldsymbol{x}}_k,\boldsymbol{s}_k)f(\boldsymbol{z}_k^j|\breve{\boldsymbol{x}}_k,\boldsymbol{s}_k) 
        & \breve{\epsilon}_k = 1,~d_k^j=0, \\
        c(\boldsymbol{z_k^j}) 
        & \breve{\epsilon}_k = 0,~d_k^j=0,\\
        1 
        & \breve{\epsilon}_k = 0 ,~d_k^j \neq 0,\\
        0 & \breve{\epsilon}_k=1 ,~d_k^j \neq 0,
    \end{cases}
\end{align}
    and the likelihood function for the detected landmarks, denoted by $t(\boldsymbol{z}_k^{c_k^i} | \boldsymbol{s}_k, \tilde{\boldsymbol{x}}_k^i,\tilde{\epsilon}_k^i, c_k^i)$, given by
\begin{align}
    t(\boldsymbol{z}_k^{c_k^i} | \boldsymbol{s}_k, \tilde{\boldsymbol{x}}_k^i,\tilde{\epsilon}_k^i, c_k^i)
    \begin{cases}
        p_\mathrm{D}(\tilde{\boldsymbol{x}}_k^i,\boldsymbol{s}_k)f(\boldsymbol{z}_k^j|\tilde{\boldsymbol{x}}_k^i,\boldsymbol{s}_k) 
        & \tilde{\epsilon}_k=1,~c_k^i=j,\\
        1-p_\mathrm{D}(\tilde{\boldsymbol{x}}_k^i,\boldsymbol{s}_k) & \tilde{\epsilon}_k=1,~c_k^i = 0,\\
        1 
        & \tilde{\epsilon}_k=0,~c_k^i =0,\\
        0 & \text{otherwise}.
    \end{cases}
\end{align}

    Now, we introduce the joint posterior density $f(\boldsymbol{s}_{k},\boldsymbol{y}_{k},\boldsymbol{c}_k,\boldsymbol{d}_k|\mathcal{Z}_{1:k})$, which can be factorized as
\begin{align}
    f(\boldsymbol{s}_{k},\boldsymbol{y}_{k}, \boldsymbol{c}_k,\boldsymbol{d}_k|\mathcal{Z}_{1:k})
    & \propto  f_{k|k-1}(\boldsymbol{s}_{k}) \prod_{i=1}^{I_{k-1}}
    f_{k|k-1}(\tilde{\boldsymbol{x}}_{k}^i,\tilde{\epsilon}_{k}^i)
    t(\boldsymbol{z}_{k}^{c_{k}^i} | \boldsymbol{s}_{k}, 
    \tilde{\boldsymbol{x}}_{k}^i,\tilde{\epsilon}_{k}^i, c_{k}^i)
    \notag \\
    & 
    \times
    \prod_{j=1}^{J_{k}} \tilde{l}(\boldsymbol{z}_{k}^j | \boldsymbol{s}_{k}, \breve{\boldsymbol{x}}_{k},\breve{\epsilon}_{k}, d_{k}^j) \psi(c_k^i, d_k^j),
    \label{eq:Ch3_BPSLAM_FactDen}
\end{align}
    where $f_{k|k-1}(\boldsymbol{s}_{k})$ denotes the predicted density of \ac{ue}, and $f_{k|k-1}(\tilde{\boldsymbol{x}}_{k}^i,\tilde{\epsilon}_{k}^i)$ denotes the predicted density of detected landmark $i$.
    
    The factorized density can be depicted by a factor graph~\cite{Frank_SPA_TIT2001}, which represents the relation among the random variables and real-valued functions for SLAM.
    The SLAM process consists of two parts: i) prediction and ii) update.
    
\begin{figure}[t!]
    \centering
    \includegraphics[width=0.8\textwidth]{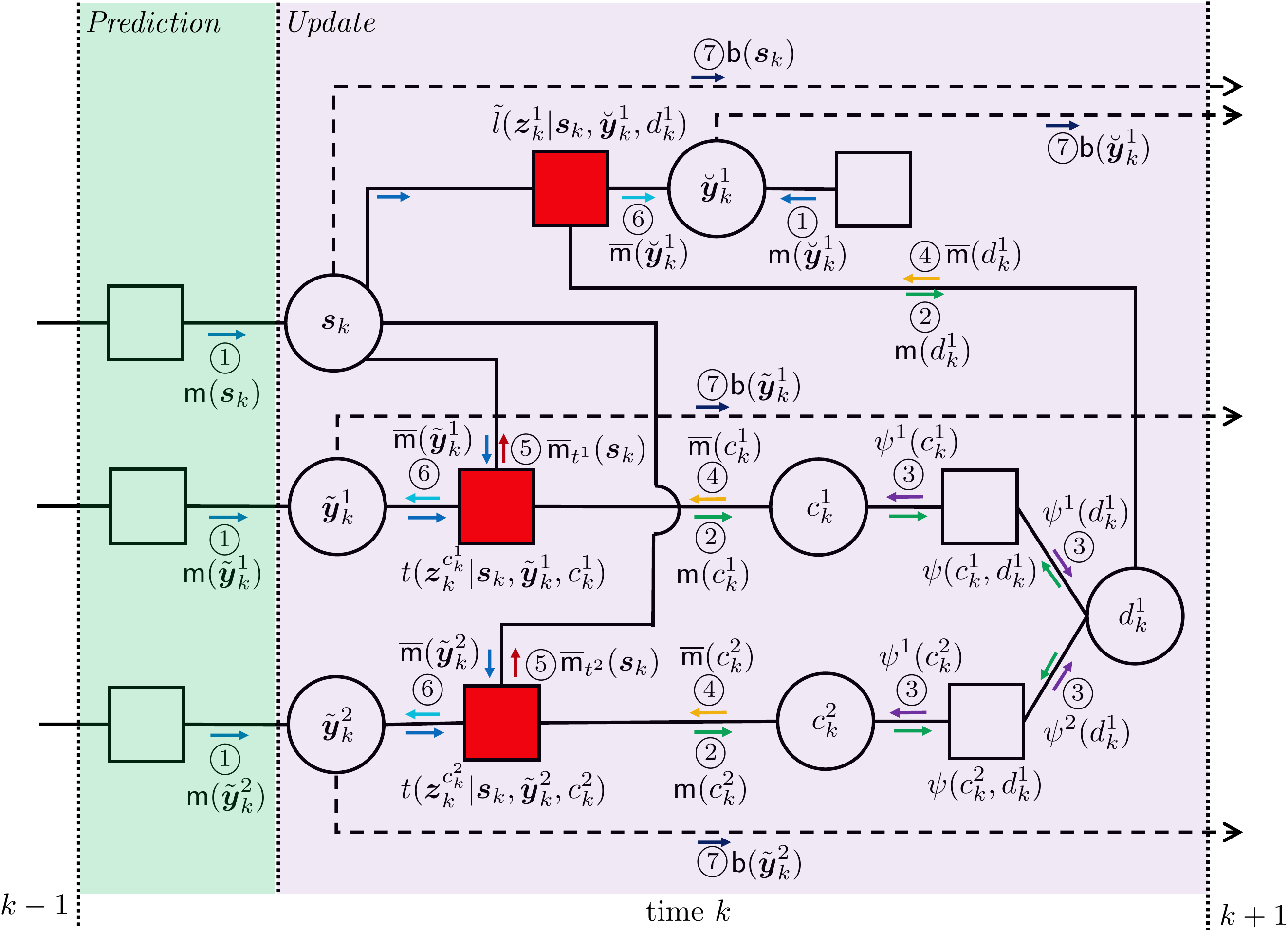}
    \caption{Factor graph representation of the joint posterior density for SLAM. Each red factor indicates the likelihood function, depending on all the measurements.}
    \label{fig:Ch3_FG}\vspace{-4mm}
\end{figure}

    For the sake of argument, let us consider a simple case as follows:
    at time $k$, two previously detected landmarks $\tilde{\boldsymbol{y}}_{k-1}^1$ and $\tilde{\boldsymbol{y}}_{k-1}^2$ are given with the \acp{PDF} $f(\tilde{\boldsymbol{x}}_{k-1}^1, \tilde{\epsilon}_{k-1}^1|\boldsymbol{z}_{1:k-1})$ and $f(\tilde{\boldsymbol{x}}_{k-1}^2, \tilde{\epsilon}_{k-1}^2|\boldsymbol{z}_{1:k-1})$; the \ac{ue} observes one measurement $\boldsymbol{z}_k^1$ (so that at least one previously detected landmark is misdetected); and the \ac{ue} prior $f(\boldsymbol{s}_{k-1}|\boldsymbol{z}_{1:k-1})$ is given.
    Then, we depict a factor graph of the simple case in Fig.~\ref{fig:Ch3_FG}, and we define the \acp{MPD} as the beliefs. For example: $f(\tilde{\boldsymbol{x}}_{k-1}^1, \tilde{\epsilon}_{k-1}^1|\boldsymbol{z}_{1:k-1}) \triangleq \mathsf{b}(\tilde{\boldsymbol{x}}_{k-1}^1, \tilde{\epsilon}_{k-1}^1)$ and $f(\boldsymbol{s}_{k-1}|\boldsymbol{z}_{1:k-1}) \triangleq \mathsf{b}(\boldsymbol{s}_{k-1})$.
    We will detail the prediction and update steps with \ac{BP} in the following section.

\subsubsection{\ac{BP} for SLAM}

    Here, we address the messages and beliefs, computed by \ac{BP} running on the factor graph.
    The messages and beliefs are interpreted with the message scheduling \large\protect\textcircled{1}--\large\protect\textcircled{7}\normalsize, as shown in Fig.~\ref{fig:Ch3_FG}.
    
\runinhead{Prediction~(Message \large\protect\textcircled{1})}

    With the transition density, the messages for sensor and previously detected landmarks are computed (see green area in Fig.~\ref{fig:Ch3_FG}).
    For example, the message for previously detected landmark 1 is 
\begin{align}
    \mathsf{m}(\tilde{\boldsymbol{y}}_k^1) = &\int f(\tilde{\boldsymbol{y}}_k^1|\tilde{\boldsymbol{y}}_{k-1}^1)\mathsf{b}(\tilde{\boldsymbol{y}}_{k-1}^1) \text{d}{\tilde{\boldsymbol{y}}}_{k-1}^1.
\end{align}
    The messages for newly detected landmarks are generated, set to the intensity function $\lambda(\breve{\boldsymbol{y}})$ of~\eqref{PPP} (shown in the pink area in Fig.~\ref{fig:Ch3_FG}). The intensity function can be represented by the \ac{GM}, and this modification outside the BP framework allows for handling the newly detected landmarks, which resembles to the process for undetected landmarks of the \ac{PMBM} filter~(see Sect.~\ref{sec:Ch3_PMBM}).
    The \ac{ue} message is $\mathsf{m}(\boldsymbol{s}_k) = \int f(\boldsymbol{s}_k|\boldsymbol{s}_{k-1})\mathsf{b}(\boldsymbol{s}_{k-1}) \text{d}\boldsymbol{s}_{k-1}$, where $f(\boldsymbol{s}_k|\boldsymbol{s}_{k-1})$ is the transition density of the sensor.


\runinhead{\ac{DA} (Message \large\protect\textcircled{2}--\large\protect\textcircled{4})}

    The messages \large\protect\textcircled{1}\normalsize~are sent to the linked factor, and then the messages \large\protect\textcircled{2}\normalsize~for landmark-oriented associations and measurement-oriented associations are obtained.
    For example, the message $\mathsf{m}(c_k^1 = 1)$ indicates detection of previously landmark 1 associated with the measurement $\boldsymbol{z}_k^1$, given by
\begin{align}
    \mathsf{m}(c_k^1 = 1) = \iint \mathsf{m}(\boldsymbol{s}_k)  \mathsf{m}(\tilde{\boldsymbol{x}}_k^1,\tilde{\epsilon}_k^1=1) 
    p_\mathrm{D}
    f(\boldsymbol{z}_k^{1} | \boldsymbol{s}_k, \tilde{\boldsymbol{x}}_k^1)
    \text{d}\boldsymbol{s}_k \text{d} \tilde{\boldsymbol{x}}_k^1,
\end{align}
    and message $\mathsf{m}(c_k^1 = 0)$ indicates missed detection of previously landmark 1, given by
\begin{align}
    \mathsf{m}(c_k^1 = 0) = \iint \mathsf{m}(\boldsymbol{s}_k)  
    \{ \mathsf{m}(\tilde{\boldsymbol{x}}_k^1,\tilde{\epsilon}_k^1=1) 
    (1-p_\mathrm{D}) + \mathsf{m}(\tilde{\boldsymbol{x}}_k^1,\tilde{\epsilon}_k^1=0) \}
    \text{d}\boldsymbol{s}_k \text{d} \tilde{\boldsymbol{x}}_k^1.
\end{align}
    The messages $\mathsf{m}(d_k^1 = 0)$ and $\mathsf{m}(d_k^1 \neq 0)$ are defined similarly. 
    To obtain the messages \large\protect\textcircled{3}\normalsize, loopy \ac{BP} is iteratively performed, detailed in~\cite{williams2015marginal}.
    The messages \large\protect\textcircled{4}\normalsize~are obtained by the product of messages \large\protect\textcircled{3}\normalsize~from the factor $\psi(\cdot,\cdot)$ to the linked variables. 
    For example, $\overline{\mathsf{m}}(d_k^1) = \psi^1(d_k^1)\psi^2(d_k^1)$.

\runinhead{Measurement Update (Message \large\protect\textcircled{5}--\large\protect\textcircled{7})}
    The messages \large\protect\textcircled{4}\normalsize~are sent to the linked factors, and the messages \large\protect\textcircled{5}\normalsize~and \large\protect\textcircled{6}\normalsize~are computed with the messages \large\protect\textcircled{1}\normalsize.
    For example, the message $\overline{\mathsf{m}}_{t^1}(\boldsymbol{s}_k)$ is computed as
\begin{align}
     & \overline{\mathsf{m}}_{t^1}(\boldsymbol{s}_k)\nonumber \\
    & =
    \int  \overline{\mathsf{m}}(c_k^1=0) (1-\mathsf{p}_\mathrm{D})\mathsf{m}(\tilde{\boldsymbol{x}}_k^1,\tilde{\epsilon}_k^1=1) +\overline{\mathsf{m}}(c_k^1 = 0) \mathsf{m}(\tilde{\boldsymbol{x}}_k^1,\tilde{\epsilon}_k^1=0)\notag \\
    &+ \overline{\mathsf{m}}(c_k^1=1) \mathsf{p}_\mathrm{D}f(\boldsymbol{z}_k^j|\tilde{\boldsymbol{x}}_k^1,\boldsymbol{s}_k)\mathsf{m}(\tilde{\boldsymbol{x}}_k^1,\tilde{\epsilon}_k^1=1) \text{d} \tilde{\boldsymbol{x}}_k^1.
\end{align}
    For example, the message $\overline{\mathsf{m}}(\tilde{\boldsymbol{x}}_k^1,\tilde{\epsilon}_k^1=1)$ and $\overline{\mathsf{m}}(\tilde{\boldsymbol{x}}_k^1,\tilde{\epsilon}_k^1=0)$ are computed as
\begin{align}
    \overline{\mathsf{m}}(\tilde{\boldsymbol{x}}_k^1,\tilde{\epsilon}_k^1=1)
    & =
    \int \overline{\mathsf{m}}(c_k^i=1) p_\textrm{D} f(\boldsymbol{z}_k^1 | \boldsymbol{s}_k, \tilde{\boldsymbol{x}}_k^1)
    \mathsf{m}(\boldsymbol{s}_k) \text{d}\boldsymbol{s}_k+ \overline{\mathsf{m}}(c_k^i=0) (1-p_\textrm{D}).\\
    \overline{\mathsf{m}}(\tilde{\boldsymbol{x}}_k^1,\tilde{\epsilon}_k^1=0)
    & = \overline{\mathsf{m}}(c_k^1 = 0).
\end{align}
    The messages \large\protect\textcircled{5}\normalsize~and \large\protect\textcircled{6}\normalsize~are sent to the linked variables.
    Finally, the beliefs (i.e., messages \large\protect\textcircled{7}\normalsize) are computed by the product of the linked factors.
    For example, the beliefs of previously detected landmark 1 ($\mathsf{b}(\tilde{\boldsymbol{x}}_k^1,\tilde{\epsilon}_k^1=1)$ and $\mathsf{b}(\tilde{\boldsymbol{x}}_k^1,\tilde{\epsilon}_k^1=0)$) are computed as
\begin{align}
    \mathsf{b}(\tilde{\boldsymbol{x}}_k^1,\tilde{\epsilon}_k^1=1) 
    &= \overline{\mathsf{m}}(\tilde{\boldsymbol{x}}_k^1,\tilde{\epsilon}_k^1=1) {\mathsf{m}}(\tilde{\boldsymbol{x}}_k^1,\tilde{\epsilon}_k^1=1), \\
    \mathsf{b}(\tilde{\boldsymbol{x}}_k^1,\tilde{\epsilon}_k^1=0) 
    &= \overline{\mathsf{m}}(\tilde{\boldsymbol{x}}_k^1,\tilde{\epsilon}_k^1=0) {\mathsf{m}}(\tilde{\boldsymbol{x}}_k^1,\tilde{\epsilon}_k^1=0).
\end{align}
    The belief of the sensor state, i.e., $\mathsf{b}(\boldsymbol{s}_k)$, is computed by the product of incoming messages from linked factors $t(\cdot)$ and $f(\boldsymbol{s}_k|\boldsymbol{s}_{k-1})$, as $    \mathsf{b}(\boldsymbol{s}_k) \propto \mathsf{m}(\boldsymbol{s}_k) \overline{\mathsf{m}}_{t^1}(\boldsymbol{s}_k)\overline{\mathsf{m}}_{t^2}(\boldsymbol{s}_k).$

\section{Results} \label{sec:results}

In this section, performance of the two RFS-SLAM algorithms is demonstrated utilizing a small cell vehicular scenario with one \ac{bs}. First, the simulation scenario and performance metrics are introduced and thereafter, the results are presented. 

\subsection{Simulation Setup}

\begin{figure}
\centerline{\includegraphics[width=0.75\linewidth]{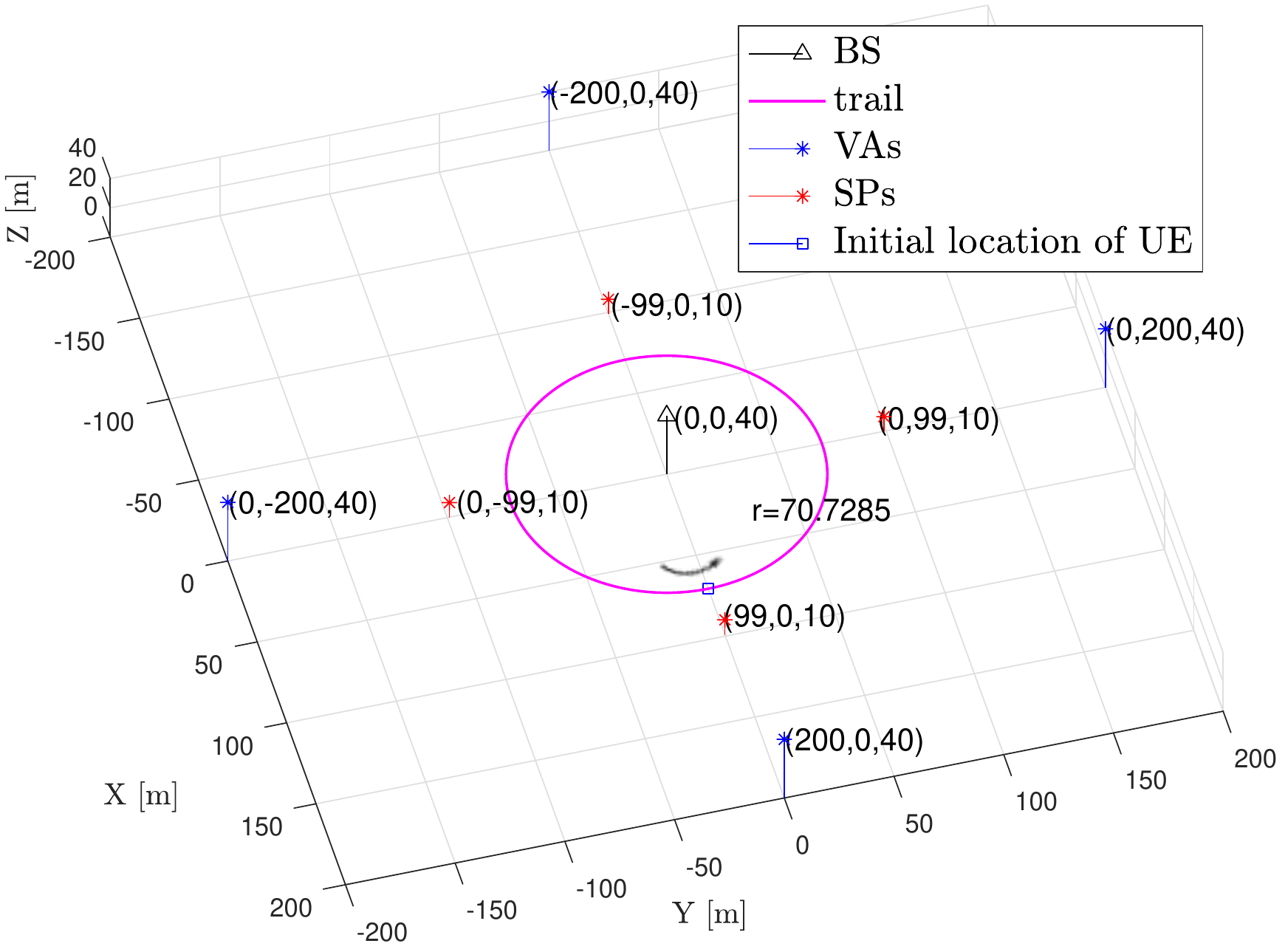}}
\caption{Scenario with the environment of a \ac{bs} and 4 VAs and 4 SPs. The \ac{ue} moves counterclockwise along the trail centered at the \ac{bs}.}
\label{fig:env}
\vspace{-4mm} \end{figure}



We consider a scenario illustrated in Fig.~\ref{fig:env} in which a \ac{ue} performs a $360^\circ$ cycle around a \ac{bs} that has a known location. The \ac{ue} state is composed of the 2D location $[x,\;y]^\top$, heading $\alpha$ and clock bias $B$, and the motion is modeled using a coordinated turn model \cite{roth2014} and the bias using a Gaussian random walk model \cite{sarkka2013}. Thus, evolution of the \ac{ue} state can be modeled as\footnote{It is to be noted that even though the complete \ac{ue} state consists of the 3D position, 3D orientation and clock parameters, a lower dimensional approximation may be sufficient. We have assumed that the terrain within a small cell is flat, the clock drift is small, and orientation and position of the antenna are known and fixed with respect to the \ac{ue} coordinate frame.}
\begin{equation}\label{dynamic_model}
     \underset{\boldsymbol{s}_{k}}{\underbrace{\vphantom{\begin{bmatrix} x_{k-1} + \tfrac{2v}{\phi}  \sin \! \left(\tfrac{\phi  \Delta}{2}\right)  \cos \! \left(\alpha_{k-1} + \tfrac{\phi \Delta}{2}\right) \\ y_{k-1} + \tfrac{2v}{\phi}\sin \! \left(\tfrac{\phi \Delta}{2}\right) \sin \! \left(\alpha_{k-1} + \tfrac{\phi \Delta}{2}\right) \\ \alpha_{k-1} + \phi \Delta \\ B_{k-1} \end{bmatrix}} \begin{bmatrix} x_{k} \\ y_{k} \\ \alpha_{k} \\ B_{k} \end{bmatrix}}} = \underset{\boldsymbol{v}\left(\boldsymbol{s}_{k-1},\boldsymbol{u}_{k} \right)}{\underbrace{\begin{bmatrix} x_{k-1} + \tfrac{2v}{\phi}  \sin \! \left(\tfrac{\phi  \Delta}{2}\right)  \cos \! \left(\alpha_{k-1} + \tfrac{\phi \Delta}{2}\right) \\ y_{k-1} + \tfrac{2v}{\phi}\sin \! \left(\tfrac{\phi \Delta}{2}\right) \sin \! \left(\alpha_{k-1} + \tfrac{\phi \Delta}{2}\right) \\ \alpha_{k-1} + \phi \delta \\ B_{k-1} \end{bmatrix}}},
\end{equation}
where $\Delta = 0.5 \text{ s}$ denotes the sampling time interval and $\boldsymbol{u}_k = [v \; \phi]^\top$ is the control input with speed $v = 22.22 \text{ m/s}$ and turn rate $\phi = \pi/10 \text{ rad/s}$. The process noise covariance is $\boldsymbol{Q} = \textrm{diag}(0.2^2 \text{ m}^2, \; 0.2^2 \text{ m}^2, \; 0.0035^2 \text{ rad}^2, \; (0.2/c)^2 \text{ ns}^2)$ where $c$ denotes the speed of light. The RFS-SLAM density is approximated using $2000$ particles and the particles are initialized using $\boldsymbol{s}_0^{(n)} \sim \mathcal{N}\left(\boldsymbol{s}_0, \boldsymbol{P}_0 \right)$, where 
$\boldsymbol{s}_0 = [v/\omega \text{ m}, \, 0 \text{ m}, \, \pi/2 \text{ rad}, \, 300/c \text{ ns}]^\top$ and
$\boldsymbol{P}_0 = \textrm{diag}(0.3^2 \text{ m}^2, \, 0.3^2 \text{ m}^2, \, 0.0052^2 \text{ rad}^2, \,\\  (0.3/c)^2 \text{ ns}^2)$. In practice, good initials can be acquired by using external sensors or by applying snapshot-based  localization algorithms, like \cite{wen20205g,kakkavas2021power}.

The landmark states are unknown and there are four \acp{va} and four \acp{sp} in the environment (see Fig.~\ref{fig:env}). The \acp{va} represent large reflecting surfaces such as walls, whereas the \acp{sp} describe small scattering objects located near the walls such as traffic signs or street lamps. The \ac{bs} and \acp{va} are always visible, whereas the \acp{sp} are only visible when in the \ac{fov} of the \ac{ue} and the \ac{fov} radius for the \acp{sp} is set to $50 \text{ m}$. The detection probability for the landmarks is set to $0.9$ if visible and $0$ otherwise. The received measurements are corrupted by additive zero-mean Gaussian noise with covariance $\boldsymbol{R} = \textrm{blkdiag}((0.1/c)^2 \text{ ns}^2, \,  10^{-4} \cdot \boldsymbol{I}_4 \text{ rad}^2)$. The initial birth intensity is ${\lambda}_{0}(\boldsymbol{x})= 1.5\times10^{-5}$. The clutter intensity is $c(\boldsymbol{z}) = 1/(4 \times 200 \pi^4)$, where the value in the numerator represents the average number of clutter measurements in the environment and $200$ in the denominator is the maximum sensing range.\footnote{The experimental setting and used parameters are the same or very close to the ones used in previous works \cite{kim2020a,kim2020b,kim20205g,ge2021computationally,ge2021iterated,EKPHD2021Ossi}. However, we want to note that our preliminary simulations indicate that the presented filters are able to cope much harder scenarios, for example when the clutter intensity is $20$ times higher. For comparative reasons, we have opted to utilize the simulation scenario used in previous works and in future research, more realistic and challenging scenarios will be considered.} Computational complexity of the filters are reduced by utilizing a Gaussian component reduction algorithm (see e.g., \cite{vo2006}) for which the pruning and  merging thresholds are set to $10^{-4}$ and $50$, respectively. 


The performance of the mapping algorithm is evaluated with the \ac{GOSPA} metric which captures the localization accuracy of the estimator and penalizes for missed and false landmarks \cite{rahmathullah2017}. The \ac{ue} state estimation accuracy is evaluated using the \ac{RMSE} and the performance is benchmarked with respect to the \ac{PCRB}. Overall, $10$ \ac{MC} simulations are performed and the results are obtained by averaging over the different rounds.

\subsection{Simulation Results}

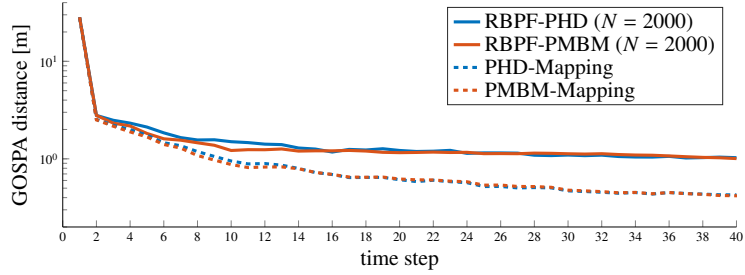
\begin{figure}
\center
\input{Mapping_VA}
\caption{Comparison of mapping performances for VAs among PHD and PMBM algorithms. The dashed lines are mapping with known poses.}
\label{fig:mapping_VA}
\vspace{-4mm} \end{figure}
\begin{figure}
\center
\input{Mapping_SP}
\caption{Comparison of mapping performances for SPs among PHD and PMBM algorithms. The dashed lines are mapping with known poses.}
\label{fig:mapping_SP}
\vspace{-4mm} \end{figure}
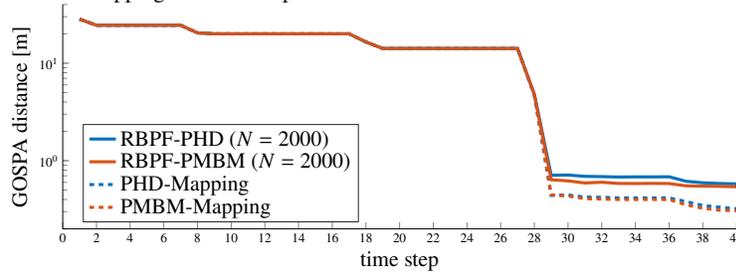

Mapping performance of the \ac{RFS}-\ac{SLAM} filters is illustrated in Fig.~\ref{fig:mapping_VA} for \acp{va} and Fig.~\ref{fig:mapping_SP} for \acp{sp}. As shown, the mapping accuracy of both filters improves gradually over time as more measurements are received. The large drops in Fig.~\ref{fig:mapping_SP} correspond to time instances the \acp{sp} are inside the \ac{fov} for the first time. At these time instances the penalization term caused by miss detection reduces by a factor $\sqrt{20^2/2} \approx 14.14$. The RBP-PMBM SLAM filter provides slightly better mapping performance than the RBP-PHD SLAM filter, as the red solid lines are slightly lower than the blue solid lines in both  Fig.~\ref{fig:mapping_VA} and Fig.~\ref{fig:mapping_SP}. The main reason the PHD filter cannot enumerate all possible \acp{DA} explicitly, which brings additional errors, e.g., some landmarks are not updated by correct measurements, due to wrong \acp{DA} or clutter, while the PMBM filter considers all \acp{DA}, so it is more stable to misdetections and clutter.


\begin{figure}
\center
\input{pos_bar}
\caption{Comparison of state estimation accuracy.}
\label{fig:ue_state}
\vspace{-4mm} \end{figure}
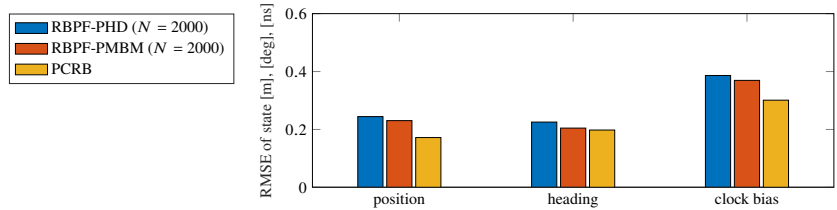

Estimation accuracy of the \ac{ue} pose and the \ac{PCRB} \cite{EKPHD2021Ossi} are summarized in Fig.~\ref{fig:ue_state}. 
The RBP-PMBM SLAM filter outperforms the RBP-PHD SLAM filter slightly, as its RMSEs are closer to the bounds. It is because that the RBP-PMBM SLAM filter explicitly considers all possible \acp{DA}, resulting in larger \ac{ESS} than the RBP-PHD SLAM. Specifically, the RBP-PMBM SLAM filter has a 6.79~\% \ac{ESS}, while the RBP-PHD SLAM filter has a 4.65~\% \ac{ESS}. However, enumerating \acp{DA} makes  the RBP-PMBM SLAM filter has much more complexity than the RBP-PHD SLAM filter, as described in Sect.~\ref{PHDMapRepresentation} and Sect.~\ref{PMBMMapRepresentation}. The accuracy of both methods could be improved using more particles but this comes with the expense of added computational complexity. 

It is valuable to mention that only mapping is required in some applications, for example when the sensor state is known or can be acquired and only the unknown environment needs to be estimated. This mapping problem is equivalent to the \ac{SLAM} problem with knowing sensor states. Pure mapping performances of the PHD and the PMBM filters are shown as dashed lines in Fig.~\ref{fig:mapping_VA} and Fig.~\ref{fig:mapping_SP}. They have the similar tendency as the mapping performances of the SLAM filters.

\subsection{Monostatic Sensing}

\Ac{ISAC} allows monostatic radar-like mapping of objects in the environment \cite{talvitie2021,barneto2022}, where a \ac{ue} with known dynamics exploits the single-bounce backscattered uplink communication signals to map the surrounding environment. This is demonstrated in the following using experimental RF measurements. The reader is referred to \cite{barneto2022} for further details on the evaluation scenario, experimental setup and channel estimator, and to \cite{talvitie2021} for details on the \ac{PHD} filter based mapping algorithm.

\begin{figure*}[!t]
\centering
\subfloat[Experimental environment]{\includegraphics[width=1.7in,trim={2.0in 0.0in 2.0in 0.0in},clip,valign=t]{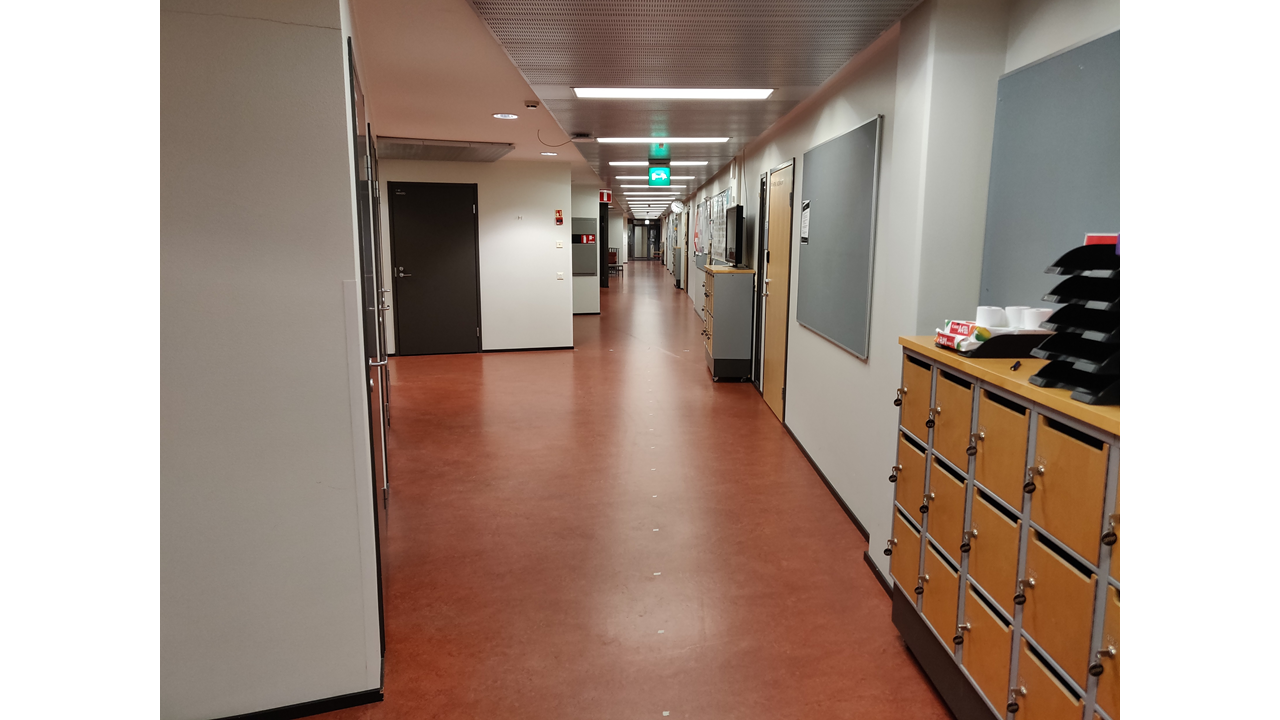}%
\vphantom{\includegraphics[width=2.9in,trim={0.2in 0.5in 0.7in 1.2in},clip,valign=t]{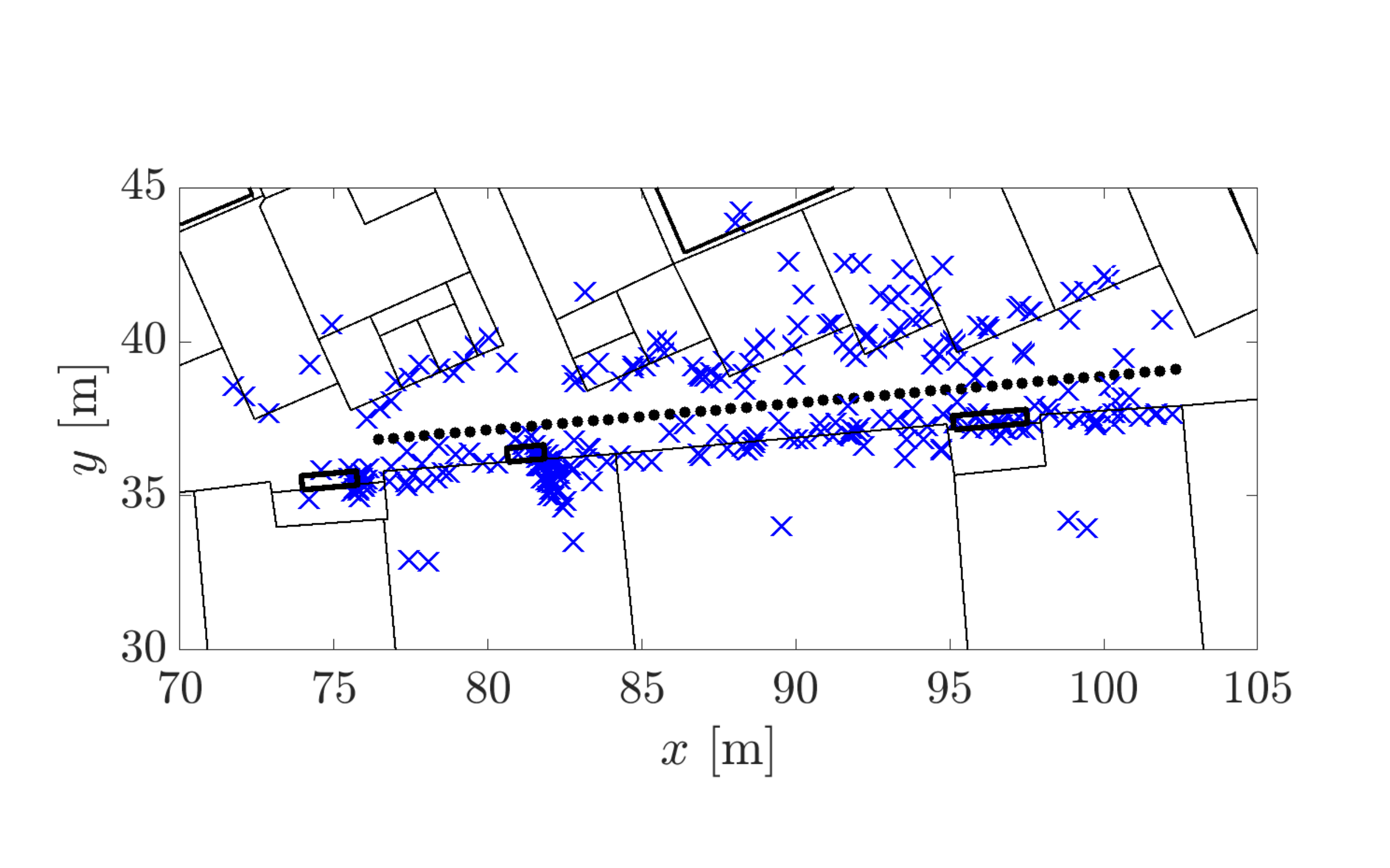}}%
\label{fig:experimental_environment}}
\hfil
\subfloat[Measurements]{\includegraphics[width=2.9in,trim={0.2in 0.5in 0.7in 1.2in},clip,valign=t]{measurements.pdf}%
\label{fig:monostatic_measurements}}
\\
\subfloat[Measurement setup]{\includegraphics[width=1.7in,trim={2.0in 0.0in 2.0in 0.0in},clip,valign=t]{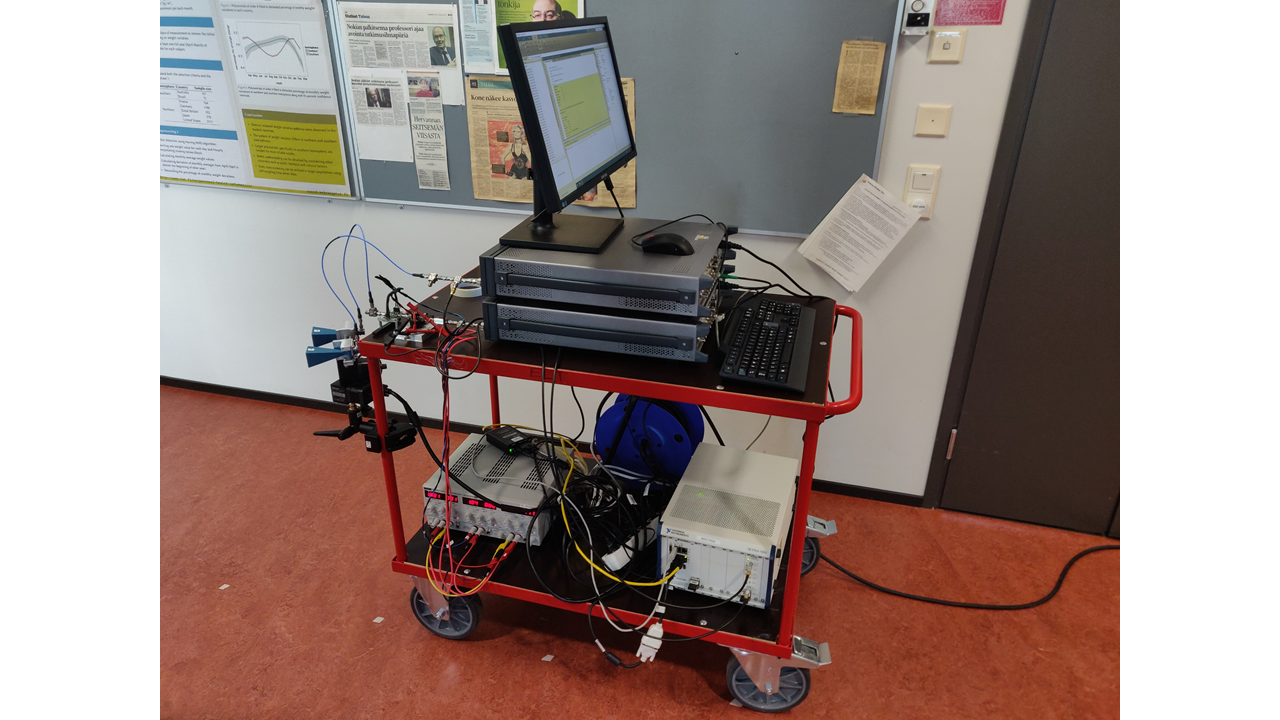}%
\vphantom{\includegraphics[width=2.9in,trim={0.2in 0.5in 0.7in 1.2in},clip,valign=t]{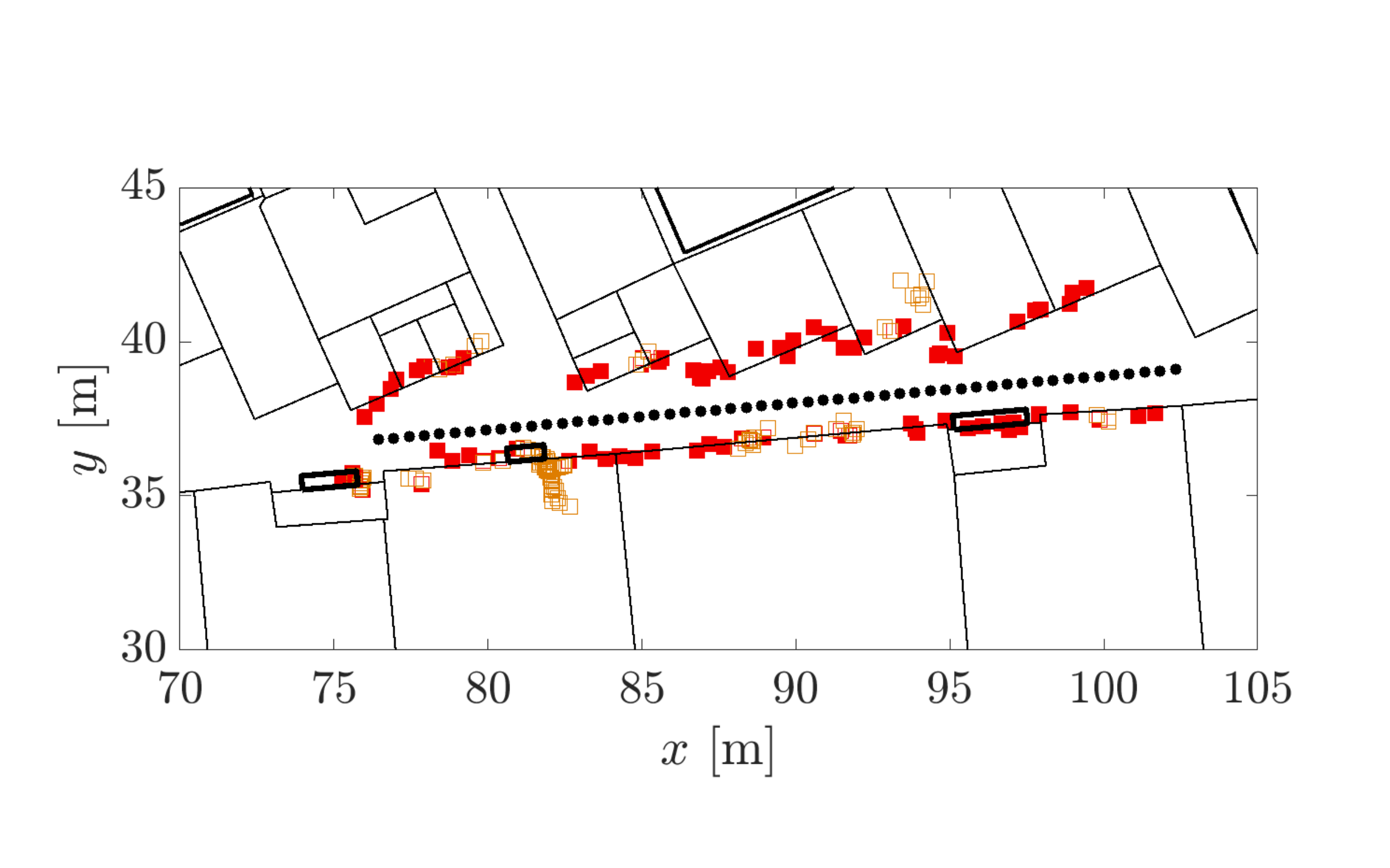}}%
\label{fig:experimental_setup}}
\hfil
\subfloat[Estimates]{\includegraphics[width=2.9in,trim={0.2in 0.5in 0.7in 1.2in},clip,valign=t]{estimates.pdf}%
\label{fig:monostatic_estimates}}
\caption{The experimental environment and measurement setup illustrated in (a) and (c), respectively. In (b), measurement acquisition locations (\protect\blackcircle) and \acf{ISTA} range-angle estimates converted to 2D Euclidean coordinates (\protect\bluecross). In (d), reflection (\protect\redsquare) and scattering (\protect\orangesquare) point estimates obtained using the \ac{PHD} filter. In (b) and (d), the black thin lines illustrate walls, whereas the black thick lines indicate wooden/metallic drawers (see (a)).}
\label{fig:monostatic_mapping} \vspace{-4mm}
\end{figure*}

The evaluation scenario is illustrated in Fig.~\ref{fig:experimental_environment}, which is a $2$ meters wide and $60$ meters long office corridor at the Hervanta Campus of Tampere University, Finland. The mapping related measurements are acquired over a $26$ meter trajectory in half a meter intervals as superimposed in Fig.~\ref{fig:monostatic_measurements}. The path traverses from right to left and the angular scanning range in each position is from  $-90^\circ$ to $90^\circ$. The RF measurements are acquired using state-of-the-art mmWave equipment shown in Fig.~\ref{fig:experimental_setup}. In the experiment, the phased array beam-steering operation is emulated using two directive horn antennas mounted on a mechanical steering system enabling accurate beam steering in the whole azimuth plane. The antennas are assumed co-located so that the system can be characterized to operate in a monostatic radar like fashion.  

The \ac{OFDM} uplink waveform follows the available 5G NR numerology at the $28$ GHz carrier frequency and the entire $400$ MHz channel bandwidth is utilized to maximize the sensing resolution. The consecutive \ac{OFDM} symbols are coherently combined to improve the SNR of the range-angle charts and the sparse map recovery problem is solved using the \ac{ISTA}. The obtained sparse range-angle charts are then subject to a target detection phase that provides the measurement input $\mathcal{Z}_k$ for the subsequent mapping algorithm. The environment is characterized by reflection and scattering landmarks for which the transition densities are known under monostatic operation and deterministic \ac{ue} movement. The measurements and landmarks are modeled using \acp{RFS} and in the following, results for a dynamic \ac{PHD} filter based mapping algorithm are presented.

An example mapping result is illustrated in Fig.~\ref{fig:monostatic_mapping}. Converting the \ac{ISTA} range-angle estimates to 2D Euclidean coordinates, which we refer to as the measurement map from now on, gives a rough outline of the corridor. However, the measurements do not align particularly well with the walls and multi-bounce signals induce artefacts behind the walls creating false landmarks into the map. The PHD filter improves the estimated map in three ways. First of all, the filtering density is conditioned on a sequence of measurements as opposed to the measurement map that solely relies on the individual measurements. Second, the mobility model acts as a spatial filter and some of the multi-bounce propagation paths are removed since they are not inline with the mobility model. Lastly, the PHD successfully removes spurious measurements which are modeled as clutter in the filter. It is expected that the \ac{PMBM} filter would yield similar or slightly better accuracy than the \ac{PHD} filter, but since we have not tested the \ac{PMBM} filter in the context of monostatic mapping, this is left for future research.

An alternative map representation is a grid-based map \cite{guerra2015,barneto2020}. A grid-based map divides the area into cells and, the range-angle measurements are projected to the corresponding cells which then represents occupancy in that cell. Grid-based mapping approaches provide a straightforward and robust solution to mapping. However, the accuracy is limited by the cell size, which itself is proportional to memory and processing capabilities, and grid maps are only suitable for mapping static environments. Feature-based approaches combine high accuracy with low complexity, and easily generalizes to moving objects. As we have demonstrated above, a feature-based \ac{PHD} filter is a viable option for mapping. Furthermore, the map could be enhanced even further using various post-processing techniques such as smoothing \cite{barneto2022,kaltiokallio2022}, but we have omitted to do so for brevity.

\section{Outlook} \label{sec:outlook}
The objective of this chapter was to  introduce the problem statements and state-of-the-art methods for localization, mapping and SLAM for ISAC in 5G and Beyond 5G communication systems. In particular, methods related to \acp{RFS} and Bayesian graphical models were highlighted. Performance evaluation of bistatic and monostatic sensing, as well as simulation and experimental results show the power of these methods, based only on standard communication signals.

With new technologies and new use cases in 6G, significant research challenges remain. For example, dealing with moving targets rather than static landmarks, which requires reasonable mobility models. The use of Doppler information, generally ignored in ISAC, could be highly beneficial. The extension to multistatic, distributed sensing scenarios is also challenging, considering both practical aspects (synchronization and calibration) as well as very difficult and distributed \ac{DA} problems. Distributed, cooperative processing and sensor fusion will provide a higher sense of situational awareness, but come with unique \ac{DA} and timing issues. Synchronization is in general a challenging topic, especially when using cheap devices with low-quality oscillators. 

Several radio enablers towards 6G will lead to specific problems, such as the extreme bandwidths at upper mmWave and THz spectrum, which will provide a very detailed view of the environment, where our simple point object models no longer hold, and where AI-based post-processing can play an important role, e.g., for determining gestures, classify objects, and perform non-radar type sensing. While scattering will be less pronounced at higher frequencies, specular reflections can appear as single-bounce (considered in this chapter), but also double- and triple bounces, which require careful handling in SLAM filters. Higher frequencies also will require more directional beamforming, which leads to radio resource management questions. Should one take a communication-centric view, where sensing is an add-on, or a service-centric view where communication and sensing are fundamental services, which each will be allocated appropriate slices (in time, frequency, and space)?
Another enabler is the user of reconfigurable intelligent surfaces (RISs), which can serve as location references, smart reflectors, and provide additional delay and angle measurements. 

In summary, there is no shortage and research questions that lie on the intersection of communication, signal processing, and artificial intelligence for the years to come. 

\begin{acknowledgement}
This work was supported, in part, by the European Commission through the H2020 project Hexa-X (Grant Agreement no. 101015956) and the Wallenberg AI, Autonomous Systems and Software Program (WASP) funded by Knut and Alice Wallenberg Foundation. The work was also supported by the Academy of Finland under the projects \#319994, \#338224, \#323244, and \#328214.
\end{acknowledgement}

\bibliographystyle{unsrt}
\bibliography{library}

\end{document}

%% file: acronyms.tex
\acrodef{bs}[BS]{base station}
\acrodef{ue}[UE]{user}
\acrodef{URA}[URA]{uniform rectangular array}
\acrodef{TOA}[TOA]{time-of-arrival}
\acrodef{rtt}[RTT]{round-trip-time}
\acrodef{TDOA}[TDOA]{time-difference-of-arrival}
\acrodef{AOA}[AOA]{angles-of-arrival}
\acrodef{AOD}[AOD]{angles-of-departure}
\acrodef{DA}[DA]{data association}
\acrodef{LOS}[LOS]{line-of-sight}
\acrodef{NLOS}[NLOS]{non-line-of-sight}
\acrodef{PMBM}[PMBM]{Poisson  multi-Bernoulli  mixture}
\acrodef{PMB(M)}[PMB(M)]{Poisson  multi-Bernoulli  (mixture)}
\acrodef{PMB}[PMB]{Poisson  multi-Bernoulli}
\acrodef{RFS}[RFS]{random finite set}
\acrodef{PPP}[PPP]{Poisson point process}
\acrodef{MBM}[MBM]{multi-Bernoulli  mixture}
\acrodef{MB}[MB]{multi-Bernoulli}
\acrodef{EKF}[EKF]{extended Kalman filter}
\acrodef{PDF}[PDF]{probability density function}
\acrodef{prs}[PRS]{positioning reference signal}
\acrodef{va}[VA]{virtual anchor}
\acrodef{sp}[SP]{scattering point}
\acrodef{fov}[FOV]{field-of-view}
\acrodef{EKF}[EKF]{extended Kalman filter}
\acrodef{CKF}[CKF]{cubature Kalman filter}
\acrodef{RBP}[RBP]{Rao-Blackwellized particle}
\acrodef{GOSPA}[GOSPA]{generalized optimal subpattern assignment}
\acrodef{RMSE}[RMSE]{root mean squared error}
\acrodef{PCRB}[PCRB]{posterior Cram{\'e}r-Rao bound}
\acrodef{MC}[MC]{Monte Carlo}
\acrodef{SLAM}[SLAM]{simultaneous localization and mapping}
\acrodef{BP}[BP]{belief propagation}
\acrodef{MPD}[MPD]{marginal posterior density}
\acrodef{MTT}[MTT]{multi-target tracking}
\acrodef{MPC}[MPC]{multipath component}
\acrodef{SIR}[SIR]{sequential importance resampling}
\acrodef{PF}[PF]{particle filter}
\acrodef{PHD}[PHD]{probability hypothesis density}
\acrodef{GM}[GM]{Gaussian mixture}
\acrodef{FISST}[FISST]{finite-set statistics}
\acrodef{ISTA}[ISTA]{iterative shrinkage/thresholding algorithm}
\acrodef{ISAC}[ISAC]{integrated sensing and communication}
 
\acrodef{LMB}[LMB]{labeled multi-Bernoulli}
\acrodef{GLMB}[$\delta$-GLMB]{$\delta$-generalized labeled multi-Bernoulli}

\acrodef{tomb}[TOMB/P]{track-oriented  marginal multi-Bernoulli/Poisson}
\acrodef{momb}[MOMB/P]{measurement-oriented  marginal multi-Bernoulli/Poisson}

\acrodef{IPLF}{iterated posterior linearization filter}

\acrodef{ESS}{effective sample size}

\acrodef{OFDM}[OFDM]{Orthogonal Frequency Division Multiplexing}

%% file: Mapping_VA.tex
\definecolor{mycolor1}{rgb}{0.00000,0.44700,0.74100}%
\definecolor{mycolor2}{rgb}{0.85000,0.32500,0.09800}%
\definecolor{mycolor3}{rgb}{0.00000,0.44700,0.74100}%
\definecolor{mycolor4}{rgb}{0.85000,0.32500,0.09800}%
\definecolor{mycolor5}{rgb}{0,0,0}%
%
\begin{tikzpicture}[scale=0.7\linewidth/14cm]

\begin{axis}[%
width=6.028in,
height=2.009in,
at={(1.011in,2.014in)},
scale only axis,
xmin=0,
xmax=40,
xlabel style={font=\color{white!15!black},font=\Large},
xlabel={time step},
ymin=.2,
ymax=40,
ymode=log,
ylabel style={font=\color{white!15!black},font=\Large},
ylabel={GOSPA distance [m]},
axis background/.style={fill=white},
axis x line*=bottom,
axis y line*=left,
legend style={legend cell align=left, align=left, draw=white!15!black,font=\Large}
]

%
%

%
%
\addplot [color=mycolor1, line width=2.0pt]
  table[row sep=crcr]{%
1 	 28.2842712474619\\ 
2 	 2.7880843963391\\ 
3 	 2.4853886183947\\ 
4 	 2.3228385044075\\ 
5 	 2.1175095841702\\ 
6 	 1.8470264333197\\ 
7 	 1.6507179333447\\ 
8 	 1.5607959488507\\ 
9 	 1.5685463128857\\ 
10 	 1.4986168289552\\ 
11 	 1.4669442698057\\ 
12 	 1.4143800599804\\ 
13 	 1.3989728873292\\ 
14 	 1.2911583746871\\ 
15 	 1.2597396990116\\ 
16 	 1.1734821238657\\ 
17 	 1.2490772745490\\ 
18 	 1.2320620059884\\ 
19 	 1.2690788412896\\ 
20 	 1.2183281213435\\ 
21 	 1.1893221116546\\ 
22 	 1.1976422126268\\ 
23 	 1.2215566724643\\ 
24 	 1.1376181682743\\ 
25 	 1.1506539156032\\ 
26 	 1.1493429225009\\ 
27 	 1.1440448350880\\ 
28 	 1.0885693238423\\ 
29 	 1.0794529729366\\ 
30 	 1.0929819868201\\ 
31 	 1.0777831039613\\ 
32 	 1.0904191708084\\ 
33 	 1.0558901066532\\ 
34 	 1.0437756804393\\ 
35 	 1.0420739020859\\ 
36 	 1.0599673486838\\ 
37 	 1.0200965413991\\ 
38 	 1.0287536990249\\ 
39 	 1.0411079264065\\ 
40 	 1.0274708642708\\ 
};
\addlegendentry{RBPF-PHD ($N=2000$)}


\addplot [color=mycolor2, line width=2.0pt]
  table[row sep=crcr]{%
1	28.2842712474619\\
2	2.78930141701978\\
3	2.32446139840791\\
4	2.17121558447884\\
5	1.82115646141388\\
6	1.60484954320867\\
7	1.55351387072331\\
8	1.46035089539028\\
9	1.37751905768746\\
10	1.2174285553943\\
11	1.24130215173988\\
12	1.23961317418424\\
13	1.26174815615497\\
14	1.20024233224708\\
15	1.20835125307248\\
16	1.20731518521581\\
17	1.21952889483442\\
18	1.20178233298079\\
19	1.16616119264323\\
20	1.1566872765527\\
21	1.16272576777966\\
22	1.17223481492582\\
23	1.16132267201094\\
24	1.16619586227553\\
25	1.12967464816263\\
26	1.13181875062247\\
27	1.12864229356282\\
28	1.14414990068978\\
29	1.13885957424995\\
30	1.1272093816621\\
31	1.11912758487675\\
32	1.12847769243793\\
33	1.10657424812441\\
34	1.09375093502827\\
35	1.08864909986161\\
36	1.07096172296724\\
37	1.05116368558258\\
38	1.03479449809016\\
39	1.02510499065596\\
40	1.00550151068623\\
};
\addlegendentry{RBPF-PMBM ($N=2000$)}

%
%

%
%
\addplot [color=mycolor1, dashed, line width=2.0pt]
  table[row sep=crcr]{%
1 	 28.2842712474619\\ 
2 	 2.5369817343009\\ 
3 	 2.2105169993125\\ 
4 	 1.9958571475115\\ 
5 	 1.7488038651990\\ 
6 	 1.4622227336075\\ 
7 	 1.3632399591344\\ 
8 	 1.1866575986038\\ 
9 	 1.0579905438911\\ 
10 	 0.9497277906169\\ 
11 	 0.8871659879340\\ 
12 	 0.8944741181527\\ 
13 	 0.8607156541147\\ 
14 	 0.7966781944874\\ 
15 	 0.7233338777787\\ 
16 	 0.6899121062752\\ 
17 	 0.6422430062748\\ 
18 	 0.6443343990111\\ 
19 	 0.6454782954822\\ 
20 	 0.6106208089047\\ 
21 	 0.5840309127962\\ 
22 	 0.6029961559650\\ 
23 	 0.5829301324485\\ 
24 	 0.5693789801494\\ 
25 	 0.5195480775775\\ 
26 	 0.5206985325560\\ 
27 	 0.5033221571243\\ 
28 	 0.5069799677808\\ 
29 	 0.5034731278923\\ 
30 	 0.4692172433956\\ 
31 	 0.4619457585075\\ 
32 	 0.4539876320037\\ 
33 	 0.4435373166480\\ 
34 	 0.4508523976395\\ 
35 	 0.4350975480747\\ 
36 	 0.4513722489219\\ 
37 	 0.4404879962771\\ 
38 	 0.4347689776182\\ 
39 	 0.4281674404470\\ 
40 	 0.4266966775860\\ 
};
\addlegendentry{PHD-Mapping}


\addplot [color=mycolor2, dashed, line width=2.0pt]
  table[row sep=crcr]{%
1	28.2842712474619\\
2	2.50481254671722\\
3	2.15975125445184\\
4	1.88812010057678\\
5	1.67629477564052\\
6	1.40370259187689\\
7	1.28796655045911\\
8	1.08055181725384\\
9	0.970744470052632\\
10	0.873944227239782\\
11	0.813724958198609\\
12	0.82362052697663\\
13	0.826045424427935\\
14	0.791333729500504\\
15	0.725191255395564\\
16	0.691843405200248\\
17	0.650491400530099\\
18	0.647126806395219\\
19	0.650179153249307\\
20	0.620904958015927\\
21	0.60928716789141\\
22	0.607155061816892\\
23	0.590484578596535\\
24	0.582038036479209\\
25	0.537352939195712\\
26	0.538276810420086\\
27	0.520892654380735\\
28	0.519371790384624\\
29	0.510682474624188\\
30	0.477231148820631\\
31	0.469413456645528\\
32	0.461086818167981\\
33	0.448548294685662\\
34	0.452744521390125\\
35	0.438477616500357\\
36	0.45118701372675\\
37	0.439451285586612\\
38	0.432790305804327\\
39	0.420502959232876\\
40	0.417758243162409\\
41	0.418716269311889\\
};
\addlegendentry{PMBM-Mapping}

\end{axis}
\end{tikzpicture}%

%% file: Mapping_SP.tex
\definecolor{mycolor1}{rgb}{0.00000,0.44700,0.74100}%
\definecolor{mycolor2}{rgb}{0.85000,0.32500,0.09800}%
\definecolor{mycolor3}{rgb}{0.00000,0.44700,0.74100}%
\definecolor{mycolor4}{rgb}{0.85000,0.32500,0.09800}%
\definecolor{mycolor5}{rgb}{0,0,0}%
%
\begin{tikzpicture}[scale=0.7\linewidth/14cm]

\begin{axis}[%
width=6.028in,
height=2.009in,
at={(1.011in,2.014in)},
scale only axis,
xmin=0,
xmax=40,
xlabel style={font=\color{white!15!black},font=\Large},
xlabel={time step},
ymin=.2,
ymax=40,
ymode=log,
ylabel style={font=\color{white!15!black},font=\Large},
ylabel={GOSPA distance [m]},
axis background/.style={fill=white},
axis x line*=bottom,
axis y line*=left,
legend pos=south west,
legend style={legend cell align=left, align=left, draw=white!15!black,font=\Large}
]

%
%

%
%
\addplot [color=mycolor1, line width=2.0pt]
  table[row sep=crcr]{%
1 	 28.2842712474619\\ 
2 	 24.5032659248780\\ 
3 	 24.5007242774168\\ 
4 	 24.5002324899663\\ 
5 	 24.5016789681090\\ 
6 	 24.5004539478380\\ 
7 	 24.5008964401672\\ 
8 	 20.4622646048076\\ 
9 	 20.0131509833640\\ 
10 	 20.0133611937891\\ 
11 	 20.0119881773558\\ 
12 	 20.0096977775858\\ 
13 	 20.0106708394470\\ 
14 	 20.0099860616740\\ 
15 	 20.0090333125572\\ 
16 	 20.0102991470173\\ 
17 	 20.0102132336465\\ 
18 	 16.5029537134969\\ 
19 	 14.1614384387760\\ 
20 	 14.1602550629907\\ 
21 	 14.1596906220332\\ 
22 	 14.1592780973325\\ 
23 	 14.1595515579472\\ 
24 	 14.1591670889720\\ 
25 	 14.1598427961679\\ 
26 	 14.1595138643238\\ 
27 	 14.1596278295811\\ 
28 	 4.7916627720049\\ 
29 	 0.7083736330933\\ 
30 	 0.7132392697282\\ 
31 	 0.6911818448010\\ 
32 	 0.6868713150480\\ 
33 	 0.6803538058926\\ 
34 	 0.6825095728698\\ 
35 	 0.6825372294598\\ 
36 	 0.6844574533427\\ 
37 	 0.6151479142839\\ 
38 	 0.5937439784893\\ 
39 	 0.5840694393835\\ 
40 	 0.5781101064955\\ 
};
\addlegendentry{RBPF-PHD ($N=2000$)}


\addplot [color=mycolor2, line width=2.0pt]
  table[row sep=crcr]{%
1	28.2842712474619\\
2	24.4983369705223\\
3	24.497474554312\\
4	24.4982508750791\\
5	24.4983517788561\\
6	24.4985383361492\\
7	24.4980601210073\\
8	20.4591598991138\\
9	20.0085949491436\\
10	20.0078953858571\\
11	20.0071216709901\\
12	20.0070581968672\\
13	20.0060507853061\\
14	20.0060892008266\\
15	20.0060764078242\\
16	20.006108852917\\
17	20.0060734914404\\
18	16.5000183047671\\
19	14.1567076914757\\
20	14.1557161229427\\
21	14.1553085550684\\
22	14.1549377939781\\
23	14.1544560802899\\
24	14.1543120724384\\
25	14.1541506677085\\
26	14.153628823289\\
27	14.1536752136521\\
28	4.77845308227962\\
29	0.637346236879395\\
30	0.620731969083888\\
31	0.590871847415731\\
32	0.601465042855217\\
33	0.58414965203808\\
34	0.583102338256643\\
35	0.585261146313085\\
36	0.582993160227075\\
37	0.552486171759295\\
38	0.548676691537558\\
39	0.545995469990216\\
40	0.540163089898967\\
};
\addlegendentry{RBPF-PMBM ($N=2000$)}

%
%

%
%
\addplot [color=mycolor1, dashed, line width=2.0pt]
  table[row sep=crcr]{%
1 	 28.2842712474619\\ 
2 	 24.4972147338942\\ 
3 	 24.4971305190562\\ 
4 	 24.4971305190562\\ 
5 	 24.4971305190562\\ 
6 	 24.4971305190562\\ 
7 	 24.4971305190562\\ 
8 	 20.4567400488743\\ 
9 	 20.0052556943538\\ 
10 	 20.0038445114270\\ 
11 	 20.0036907056784\\ 
12 	 20.0034232689581\\ 
13 	 20.0033467281302\\ 
14 	 20.0033467281302\\ 
15 	 20.0033467281302\\ 
16 	 20.0033467281302\\ 
17 	 20.0033467281302\\ 
18 	 16.4944974759445\\ 
19 	 14.1504934678351\\ 
20 	 14.1495941111210\\ 
21 	 14.1484730387945\\ 
22 	 14.1482379670914\\ 
23 	 14.1482918355992\\ 
24 	 14.1482918355992\\ 
25 	 14.1482918355992\\ 
26 	 14.1482918355992\\ 
27 	 14.1482918355992\\ 
28 	 4.6628463459784\\ 
29 	 0.4407485795316\\ 
30 	 0.4448858930235\\ 
31 	 0.4233156172628\\ 
32 	 0.4212663355632\\ 
33 	 0.4154613757081\\ 
34 	 0.4154613757081\\ 
35 	 0.4154613757081\\ 
36 	 0.4154613757081\\ 
37 	 0.3810083177486\\ 
38 	 0.3474558710956\\ 
39 	 0.3334000046261\\ 
40 	 0.3220074686763\\ 
};
\addlegendentry{PHD-Mapping}


\addplot [color=mycolor2, dashed, line width=2.0pt]
  table[row sep=crcr]{%
1	28.2842712474619\\
2	24.4970350243814\\
3	24.496887480537\\
4	24.496887480537\\
5	24.496887480537\\
6	24.496887480537\\
7	24.496887480537\\
8	20.4568367940661\\
9	20.005545228312\\
10	20.0037827803057\\
11	20.003527134537\\
12	20.0031846953742\\
13	20.0030818714639\\
14	20.0030818714639\\
15	20.0030818714639\\
16	20.0030818714639\\
17	20.0030818714639\\
18	16.4949763936624\\
19	14.1492121444357\\
20	14.1488111890278\\
21	14.1480376094464\\
22	14.1477693510062\\
23	14.1477412409115\\
24	14.1477412409115\\
25	14.1477412409115\\
26	14.1477412409115\\
27	14.1477412409115\\
28	4.6729687466659\\
29	0.446151696569783\\
30	0.434089687754112\\
31	0.407568428169952\\
32	0.404436475028266\\
33	0.399926570334297\\
34	0.399926570334297\\
35	0.399926570334297\\
36	0.399926570334297\\
37	0.352981651472961\\
38	0.3252377375775\\
39	0.310799861811079\\
40	0.306456092037541\\
};
\addlegendentry{PMBM-Mapping}

\end{axis}
\end{tikzpicture}%

%% file: pos_bar.tex
%
%
\definecolor{mycolor1}{rgb}{0.00000,0.44700,0.74100}%
\definecolor{mycolor2}{rgb}{0.85000,0.32500,0.09800}%
\definecolor{mycolor3}{rgb}{0.92900,0.69400,0.12500}%
\definecolor{mycolor4}{rgb}{0.49400,0.18400,0.55600}
\definecolor{mycolor5}{rgb}{0,0,0}%
\begin{tikzpicture}[scale=0.85\linewidth/14cm]

\begin{axis}[%
width=3.842in,
height=1.281in,
at={(3.465in,2.378in)},
scale only axis,
bar shift auto,
xmin=0.5,
xmax=3.5,
xtick={1,2,3},
xticklabels={{position},{heading},{clock bias}},
ymin=0,
ymax=0.6,
ylabel style={font=\color{white!15!black}},
ylabel={RMSE of state [m], [deg], [ns]},
axis background/.style={fill=white},
legend style={at={(-0.15,1)}, anchor=north east, legend cell align=left, align=left, draw=white!15!black}
]

%
%

%
%
\addplot[ybar, bar width=0.145, fill=mycolor1, draw=black, area legend] table[row sep=crcr] {%
1	0.2444 \\
2	0.2255 \\
3	0.3864 \\
};
\addplot[forget plot, color=white!15!black] table[row sep=crcr] {%
0.5	0\\
3.5	0\\
};
\addlegendentry{RBPF-PHD ($N=2000$)}

\addplot[ybar, bar width=0.145, fill=mycolor2, draw=black, area legend] table[row sep=crcr] {%
1	0.2305\\
2	0.2047\\
3	0.3695\\
};
\addplot[forget plot, color=white!15!black] table[row sep=crcr] {%
0.5	0\\
3.5	0\\
};
\addlegendentry{RBPF-PMBM ($N=2000$)}

%
%

%
%
\addplot[ybar, bar width=0.145, fill=mycolor3, draw=black, area legend] table[row sep=crcr] {%
1	0.1720\\
2	0.1980\\
3	0.3011\\
};
\addplot[forget plot, color=white!15!black] table[row sep=crcr] {%
0.5	0\\
3.5	0\\
};
\addlegendentry{PCRB}

\end{axis}
\end{tikzpicture}%